\documentclass[preprint,superscriptaddress,nobibnotes,amsmath,amssymb,aps,prmaterials,floatfix]{revtex4-2}

\usepackage{graphicx}
\usepackage{dcolumn}
\usepackage{bm}
\usepackage[colorlinks=true,urlcolor=blue,citecolor=magenta]{hyperref}

\usepackage{graphicx,threeparttable,epsfig}
\usepackage{booktabs} 
\newcommand{\ra}[1]{\renewcommand{\arraystretch}{#1}}
\usepackage{array} 

\begin{document}

\title{Thermoelectric properties of cement composite analogues from first principles calculations}

\author{Esther Orisakwe}
\email{e.orisakwe@qub.ac.uk}
\email{estherorisakwe@gmail.com}
\affiliation{School of Mathematics and Physics, Queen's University Belfast, UK}
\author{Conrad Johnston}
\affiliation{Pacific Northwest National Laboratory, Richland, WA, USA}
\author{Ruchita Jani}
\affiliation{School of Civil and Structural Engineering, Technological University Dublin, Ireland}
\author{Xiaoli Liu}
\affiliation{Lyles School of Civil Engineering, Purdue University, West Lafayette, IN, USA}
\affiliation{Multifunctional Equipment Integration Group, Oak Ridge National Laboratory, Oak Ridge, TN, USA}
\author{Lorenzo Stella}
\affiliation{School of Mathematics and Physics, Queen's University Belfast, UK}
\affiliation{School of Chemistry and Chemical Engineering, Queen's University Belfast, UK}
\author{Jorge Kohanoff}
\affiliation{Instituto de Fusion Nuclear ``Guillermo Velarde'', Universidad Politecnica de Madrid, Spain}
\affiliation{School of Mathematics and Physics, Queen's University Belfast, UK}
\author{Niall Holmes}
\affiliation{School of Civil and Structural Engineering, Technological University Dublin, Ireland}
\author{Brian Norton}
\affiliation{School of Civil and Structural Engineering, Technological University Dublin, Ireland}
\affiliation{University College Cork, Ireland}
\affiliation{Tyndall National Institute, Cork, Ireland}
\author{Ming Qu}
\affiliation{Lyles School of Civil Engineering, Purdue University, West Lafayette, IN, USA}
\author{Hongxi Yin}
\affiliation{Center for Energy, Environment \& Sustainability, Washington University in St Louis (WUST), St Louis, MO, USA}
\author{Kazuaki Yazawa}
\affiliation{Birck Nanotechnology Center, Purdue University, West Lafayette, IN, USA}

\date{\today}

\begin{abstract}
Buildings are responsible for a considerable fraction of the energy wasted globally every year, and as a result, excess carbon emissions. While heat is lost directly in colder months and climates, resulting in increased heating loads, in hot climates cooling and ventilation is required. One avenue towards improving the energy efficiency of buildings is to integrate thermoelectric devices and materials within the fabric of the building to exploit the temperature gradient between the inside and outside to do useful work. 
Cement-based materials are ubiquitous in modern buildings and present an interesting opportunity to be functionalised. We present a systematic investigation of the electronic transport coefficients relevant to the thermoelectric materials of the calcium silicate hydrate (C-S-H) gel analogue, tobermorite, using Density Functional Theory calculations with the Boltzmann transport method.
The calculated values of the Seebeck coefficient are within the typical magnitude (200 - 600 $\mu V/K$) indicative of a good thermoelectric material \cite{Guo2015}. The tobermorite models are predicted to be intrinsically $p$-type thermoelectric material because of the presence of large concentration of the Si-O tetrahedra sites. The calculated electronic $ZT$ for the tobermorite models have their optimal values of 0.983 at (400 $\mathrm{K}$ and $10^{17}$ $\mathrm{cm^{-3}}$) for tobermorite 9 \r{A}, 0.985 at (400 $\mathrm{K}$ and $10^{17}$ $\mathrm{cm^{-3}}$) for tobermorite 11 \r{A} and 1.20 at (225 $\mathrm{K}$ and $10^{19}$ $\mathrm{cm^{-3}}$) for tobermorite 14 \r{A}, respectively.
\end{abstract}

\maketitle


\section{Introduction}
On a global basis, more than 60$\%$ of the energy consumed is ultimately lost via heat to the environment\cite{Yu2009}. It was reported that buildings contribute somewhere between 20 to 40$\%$ of the globally generated energy and energy related environmental emissions, with a sizeable amount of this energy going towards heating and cooling requirements of buildings \cite{Perez-Lombard2008}. The overwhelming influence of climate change as well as the increasing energy demand in urban areas calls for more buildings to maximise their potential energy harvesting, distribution, storage and efficient usage \cite{Zhang2018,Serale2018}. Thermoelectric (TE) units can make use of the unavoidable loses to generate free electric power, because of their ability to partially recover heat by converting it into electricity \cite{Kim2014,Liu2015}.\\
In thermoelectric materials, a temperature gradient leads to a movement of charge carriers, which results into a difference of potential at the two ends of an open circuit. In this way, thermal energy can be partially converted into electrical energy with a finite efficiency ultimately bounded by the second low of the thermodynamics. A quick estimate of the energy conversion efficiency of a thermoelectric material is given by the dimensionless figure of merit, $ZT$\cite{Snyder2008}:
\begin{equation}
\label{zt}
    ZT = \frac{S^2 \sigma}{\kappa}T,  
\end{equation}
where $S$ is the Seebeck coefficient, $\sigma$ is the electrical conductivity, $T$ is the absolute temperature, and the total thermal conductivity $\kappa= \kappa_e + \kappa_l$ is made up of contributions from charge carriers $\kappa_e$ and lattice vibrations $\kappa_l$. To maximise the figure of merit, the Seebeck coefficient and electrical conductivity must be maximised, while at the same time, the thermal conductivity must be minimised\cite{Heremans2012,Han2014}. In the limit of $ZT\to\infty$, the efficiency of a Carnot engine is retrieved. For practical applications, a figure of merit $ZT\gg1$ is typically required \cite{Mahan1997}.\\
Cementitious materials are complex heterogeneous composites formed by the admixture of calcareous (mostly calcium carbonate) material such as limestone with silica-, alumina- and iron-based material, which are calcinated until they fuse together. Cement-based materials like concrete are not only used as structural materials for buildings, but also for other applications like nuclear waste disposal and massive structures like water reservoirs and dams \cite{Mehta2001,Taylor1997}.
\\
In recent years, attention has turned to composite cement-based materials. One such composite, carbon fiber reinforced cement (CFRC), has ignited great interest in the scientific community as a promising thermoelectric material\cite{Chen1993}. While standard cement formulations show only a mild thermoelectric effect \cite{Sun1998a,Sun1998b,Wen1999,Wen2000,Wen2001,Demirel2008}, the inclusion of additives such as carbon and steel fibers, inorganic compounds like graphite and/or metallic oxides such as ZnO, Bi$_2$O$_3$ or Fe$_2$O$_3$, or standard thermoelectric materials like Bi$_2$Te$_3$, can enhance the thermometric performance considerably. The challenges and opportunities presented by these composite materials have been discussed extensively in the literature \cite{Zuo2012,Wei2014,Wei2014a,Ji2016,Wei2016, Wei2017,Wei2018,Liu2020}, and in a recent review\cite{Liu2021}. While rapid progress has been made in improving the Seebeck coefficient within CFRCs, the overall figure of merit has not been improved due, understandably, to reportedly small values of electrical conductivity.
\\
The most important hydration product of cementitious materials is Calcium-Silicate-Hydrate gel, denoted C-S-H within the cement chemistry community, where C=CaO, S=SiO$_2$ and H=H$_2$O \cite{Allen2007}. C-S-H gel has an average calcium to silicon (Ca/Si) ratio of 1.7 \cite{Allen2007}, with local Ca/Si ratio fluctuations of 0.67 to 2.0 \cite{Zhang2000}. The description of the microscopic structure of cement is particularly complex due to the disordered and inhomogeneous nature of the material, which comprises several coexisting phases like alite, belite, etc \cite{Gartner2017}.  After the work of Taylor \cite{Taylor1997}, the consensus is that C-S-H based materials have a layered atomic structure akin to those of tobermorite, T14\r{A} [Ca$_5$Si$_6$O$_{16}$(OH)$_2$.7H$_2$O] and jennite [Ca$_9$Si$_6$O$_{18}$(OH)$_6$.8H$_2$O], with a calcium to silicon ratio of 0.83 and 1.5, respectively\cite{Jiang2018}. Tobermorite minerals are generally characterized by their interlayer spacing and are named accordingly as tobermorite, T9\r{A}, tobermorite, T11\r{A} and tobermorite, T14\r{A} \cite{Kumar2017}. Extensive structural characterization via XRD experiments of C-S-H based materials has been carried out by the group of Merlino and Bonaccorsi \cite{Merlino1999,Merlino2001,Bonaccorsi2005,Bonaccorsi2004}. So far the first principles density functional theory calculations carried out on the C-S-H based materials such as jennite structure and tobermorite minerals has been centered around the understanding of their structural properties \cite{Shahsavari2009,Jiang2018}, elastic constants \cite{Shahsavari2009,Vidmer2014,Jiang2018} and anisotropic effect\cite{Jiang2018}, average mechanical properties \cite{Shahsavari2009,Jiang2018}, vibrational properties and infrared spectra \cite{Vidmer2014} and/or NMR shift investigations \cite{Kumar2017}. Recently, some properties of C-S-H within the context of the civil nuclear industry have been assessed, like its ability to trap and hold radioactive fission products like Sr and its daughters \cite{Dezerald2015,Kohanoff2021}, and hydrogen gas production as a consequence of irradiation \cite{Lecaer2017}. Since the C-S-H gel is responsible for the cohesive strength and durability of cementitious materials \cite{Lothenbach2011,Lothenbach2015}, our drive is to optimize the thermoelectric transport coefficient of some C-S-H based tobermorite materials with the aim of improving their energy efficiency.\\
Motivated by the experimental results on cement-based materials, we present here first-principles calculations of the structural and electronic transport properties of the C-S-H analogs tobermorite, T9\r{A} [Ca$_5$Si$_6$O$_{16}$(OH)$_2$], tobermorite, T11\r{A} [Ca$_4$Si$_6$O$_{15}$(OH)$_2$.5H$_2$O], and tobermorite, T14\r{A} [Ca$_5$Si$_6$O$_{16}$(OH)$_2$.7H$_2$O] as shown in Fig.~\ref{fig1}. We are mainly interested in the thermoelectric properties in the low temperature regime since most of the experiments have been carried out at temperatures below 100 $^{\circ}$C. In particular, we focus on the effects of temperature and doping on electrical and thermal conductivity, Seebeck coefficient, power factor, and figure of merit. In this work, doping is modelled in terms of an intrinsic carrier concentration determined by the electronic chemical potential via the rigid band model. To the best of our knowledge, until now no calculations have been performed to understand the thermoelectric behavior of calcium silicate hydrate (C-S-H)-based materials. Our initial first-principles study can guide further investigations of the electronic transport properties of this class of materials.

 \section{Computational Methods}
%
\subsection{Details of electronic structure calculations}
All our calculations were carried out within the framework of density functional theory (DFT) \cite{Hohenberg1964,Kohn1965} using a plane wave basis set as implemented in the open-source package Quantum ESPRESSO \cite{Giannozzi2009}. All structural  optimizations and electronic property calculations were carried out using the Perdew-Burke-Ernzerhof (PBE) \cite{Perdew1996} exchange-correlation functional belonging to the generalized gradient approximation (GGA) family \cite{DalCorso1996}. Ultra-soft pseudopotentials \cite{Vanderbilt1990}
were used and along with a plane wave basis with an energy cutoff of 70 Ry and density cutoff of 560 Ry for T9\r{A} [Ca$_5$Si$_6$O$_{16}$(OH)$_2$] while energy cutoff of 80 Ry  and density cutoff 640 Ry was used for T11\r{A} [Ca$_4$Si$_6$O$_{15}$(OH)$_2$.5H$_2$O] and T14\r{A} [Ca$_5$Si$_6$O$_{16}$(OH)$_2$.7H$_2$O], respectively. The Brillouin zone was sampled using the Monkhorst-Pack scheme \cite{Pack1977}, with a regular mesh of 3$\times$3$\times$3 for T9\r{A} and T11\r{A}, and 3$\times$3$\times$1 for T14\r{A}. Electronic occupations were smeared using the Marzari-Vanderbilt scheme with broadening of $0.03$ Ry \cite{Marzari1999} and the SCF convergence threshold for the electronic wavefunctions was set to $10^{-10}$ Ry for both the variable-cell and ionic relaxation. Both jennite and tobermorite models have layered structures, with layers bound by non-bonded interactions \cite{Jiang2018}. To include the effect of these interactions and obtain realistic interlayer distances during lattice constant optimization, the semiempirical dispersion correction of Grimme \textit{et al.} \cite{Grimme2010} for the PBE functional (DFT-D3) was used.
Atomic positions and cell parameters were fully optimized at 0 K using the Broyden-Fretcher-Goldfarb-Shanno method \cite{Broyden1970,Fletcher1970,Goldfarb1970,Shanno1970}. We optimized all stress and force components to less than $0.5$ kbar and $10^{-6}$ Ry/Bohr, respectively.\\
Finally we performed a single point energy calculation in order to ensure the robustness of our results with much denser $k$-point grid of 5$\times$5$\times$5 for all the tobermorite models. Since the electronic transport coefficients are strongly dependent on the band structure energies in the Brillouin zone, we performed non-self consistent calculations for all the tobermorite models using 9$\times$9$\times$9 Monkhorst-Pack $k$-points grid (365 $k$-points in the irreducible Brillouin zone), while maintaining the energy criterion above. These non-SCF calculations were used to obtain the band structures and transport coefficients of the tobermorite models.

\subsection{Transport Calculations}
\label{subsec1}
Electronic transport coefficients were computed from the solution of the  linearized Boltzmann transport equation in conjunction with the rigid-band approximation (RBA) and constant relaxation time approximations (CRTA) \cite{Nag1980,Ash1976} as implemented in the BoltzTraP code.\cite{Madsen2006} In the rigid band approximation, it is assumed that only the chemical potential, not the band shape of the host compound, is changed according to the nominal doping and/or temperature. Hence, only the band energies are extrapolated for each calculation \cite{Scheidemantel2003}. The electronic band energies of all the tobermorite models were calculated on a 9$\times$9$\times$9 Monkhorst-Pack $k$-point grid and Fourier interpolated over a denser grid containing $15$ times as many $k$-points for better numerical evaluation of the integrals, to obtain the thermoelectric transport coefficients. In the semi-classical Boltzmann approach, the energy-dependent transport distribution function - the kernel of all transport coefficients \cite{Madsen2006,J.M.Ziman1972,Mahan1996,Takeuchi2012,Madsen2018} is expressed as:  
\begin{eqnarray}
\label{eqn1}
\Sigma_{\alpha\beta}(\epsilon)&=&\frac{q^2}{V} \sum_{ik} \tau_{ik}v_\alpha(i,k)v_\beta(i,k) \delta(\epsilon-\epsilon(i,k)),
 \end{eqnarray}
where $\alpha$ and $\beta$ are Cartesian components and the subscripts $i$ and $k$ are the band and wave vector indices, respectively. $V$ is the unit cell volume, $\epsilon(i,k)$ represents the $k$-dependent band energies, $q$ is the electron charge, $\tau_{ik}$ is the relaxation time and $v_\alpha(i,k)$ is the $\alpha$ component of the electron group velocity written as:   
\begin{eqnarray}
v_\alpha(k) &=& \frac{1}{\hbar}\frac{\partial \epsilon(i,k)}{\partial k_\alpha}.
\end{eqnarray}
The advantage of the RBA in conjunction with the CTRA, is that the transport distribution function $\Sigma_{\alpha\beta}\left(\epsilon\right)$ does not depend on either temperature $T$ or electron chemical potential (doping) $\mu$. Therefore, by integrating over a fixed transport distribution function, $\Sigma_{\alpha\beta}\left(\epsilon\right)$, the doping and/or temperature-dependent generalised transport coefficient of the $p$-th order, which is solely due to the Fermi-Dirac distribution function $f_{0}$ is obtained as:
\begin{eqnarray}
L^p(\mu, T)&=& \int\Sigma_{\alpha\beta}(\mu, T)(\epsilon(i,k)-\mu)^p\left(-\frac{\delta f_0(\epsilon,\mu,T)}{\delta\epsilon}\right)\delta\epsilon,
 \end{eqnarray}
From where the electrical conductivity $\sigma$, Seebeck coefficient $S$, and electronic thermal conductivity $\kappa_e$  are extracted as a function of temperature, $T$, and electronic chemical potential, $\mu$. These transport coefficients are written as:
\begin{eqnarray}
\sigma(\mu,T)&=&\frac{1}{V}\int\Sigma_{\alpha\beta}(\mu,T)\left(-  \frac{\partial f_0(\epsilon,\mu,T)}{\partial \epsilon}\right)d\epsilon, \end{eqnarray}
\begin{eqnarray}
\label{eqn5}
S(\mu,T)&=& \dfrac{1}{VqT\sigma(\mu,T)}\int\Sigma_{\alpha\beta}(\mu,T)(\epsilon-\mu)\left(-\frac{\partial f_0(\epsilon, \mu,T)}{\partial \epsilon}\right)d\epsilon, 
\end{eqnarray}
and    
\begin{eqnarray}
\kappa_e(\mu,T)&=&\frac{1}{VT}\int\Sigma_{\alpha\beta}(\mu,T)(\epsilon-\mu)^2\left(-\frac{\partial f_0(\epsilon,\mu,T)}{\partial \epsilon}\right)d\epsilon - TS^2\sigma
\end{eqnarray}
At zero temperature, $\mu$ is equal to the Fermi energy, $E_F$. At fixed doping, the chemical potential steadily deviates from $E_F$ with increasing temperature as a result of the energy dependence of the density of states \cite{Takeuchi2012}. The electrical conductivity is related to the electronic thermal conductivity $\kappa_e$  through the Wiedemann-Franz law, $\kappa_e=L\sigma T$, where $L$ is the Lorenz number \cite{Snyder2008}. The electrical conductivity is given as a function of $\tau(i,k)$ (Eq.~\ref{eqn1}) and, therefore, the relaxation time $\tau$ must be included as a parameter. 

To obtain $\tau$, one needs to calculate the scattering of electrons by ionized impurities, piezoelectric scattering, and/or acoustic phonon scattering by deformation \cite{Scheer2002}. From Eq.~\ref{eqn1}, it follows that $\tau \propto (\epsilon-\epsilon(i,k))^{r-1/2}$, which sometimes has been used to understand the electron transport processes but has not been established in many materials \cite{Takeuchi2012}.  For a simple parabolic band model, $\tau$ is expressed as a power-law in a reduced carrier energy $\eta=E_F/k_{B}T$, $\tau \propto \epsilon^{r-1/2}$, where $r$ is the scattering parameter\cite{Takeuchi2012,Anselm1981}. Within the rigid-band approximation, $\tau$ is typically assumed to be energy-independent and constant for simplicity \cite{Schulz1992,Madsen2006}. In this case, $\tau$ cancels out in the expression for the Seebeck coefficient, which is one of the key quantities for the thermoelectric figure of merit, $ZT$. 
The calculated transport coefficients form the essential part of the figure of merit $ZT$, which can be rearranged as:
\begin{eqnarray}
\label{eqn3}
ZT &=& \frac{S^2 \sigma~T}{\kappa_e} \left(\frac{\kappa_e}{\kappa_e+\kappa_l} \right)
\end{eqnarray}
Two contributions to $ZT$ can be identified in Eq.~(\ref{eqn3}): (i) the electronic figure of merit ($Z_eT = S^2 \sigma T/\kappa_e$), which is independent of the relaxation time and obtained when the heat transferred to the material through lattice vibrations ($\kappa_l$) is completely neglected; and (ii) a scaling factor determined by the ratio of $\kappa_e/(\kappa_e+\kappa_l)$. The electronic figure of merit $Z_eT$ approaches $ZT$ when $\kappa_l$ becomes much smaller than $\kappa_e$. Note that the final thermoelectric transport coefficients were evaluated from their averaged trace components.


\section{Results and discussion}
\subsection{Structural details of Tobermorite models}
\begin{figure}[!h]
    \centering
    (a)\includegraphics[width=0.35\textwidth]{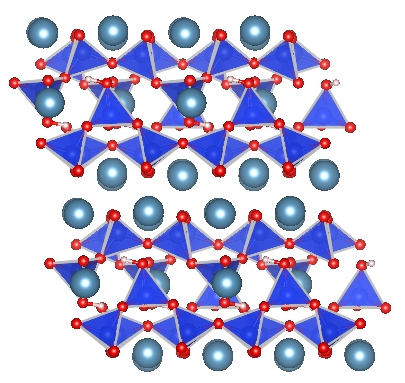}
    (b)\includegraphics[width=0.35\textwidth]{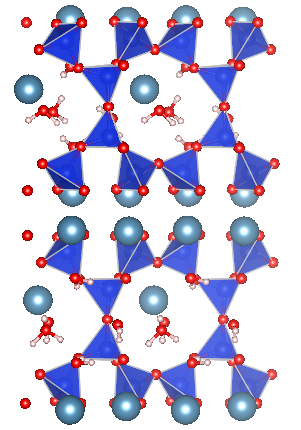}
    (c)\includegraphics[width=0.35\textwidth]{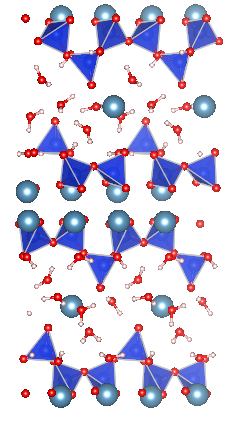}
    \caption{Layered crystal structure of tobermorite mineral: (a) T9\r{A} (Ca$_5$Si$_6$O$_{16}$(OH)$_2$), (b) T11\r{A} (Ca$_4$Si$_6$O$_{15}$(OH)$_2$.5H$_2$O)  and (c) T14\r{A} (Ca$_5$Si$_6$O$_{16}$(OH)$_2$.7H$_2$O). The large grey spheres represent calcium atoms, red spheres represent oxygen atoms, and white spheres represent the hydrogen atoms, respectively. The silicate tetrahedra are represented in blue, with the silicon atoms located in the center of the tetrahedra.}
    \label{fig1}
\end{figure}

A representation of the different tobermorite crystal structures (T9\r{A}, T11\r{A} and T14\r{A}) is shown in Fig.~\ref{fig1}. These layered tobermorite minerals share similar structural arrangements, but are distinguished by the interlayer spacing distance. The notation 9\r{A}, 11\r{A} and 14\r{A} indicates the characteristic interlayer spacings of $0.93$ nm, $1.13$ nm and $1.4$ nm. This spacing is affected by the degree of hydration, a property that changes through heating. Each layer consists of continuous seven-fold edge-sharing Ca-O polyhedra stretched along the b-direction. All the oxygen atoms of the polyhedra are shared with silicon atoms to form corner-sharing [SiO$_4$]$^{4-}$ tetrahedra chains. These chains are either dreierketten or wollastonite-like type \cite{Biagioni2011,Biagioni2015} with repeating units of three-fold tetrahedra. Two of the tetrahedra chains share edges with the Ca-polyhedron in a paired form while the third tetrahedra chain is linked to the calcium polyhedra through their apical oxygen atoms in a bridging form. Interlayer calcium polyhedra are present in all three tobermorite models. In T11\r{A} and T14\r{A} the interlayer spacing is filled with water molecules and additional calcium atoms. T11\r{A} and T14\r{A} are both different from T9\r{A} because of the formation of double (condensed) dreierketten bridging silicate chains \cite{Merlino1999,Merlino2001,Bonaccorsi2005}. 
Single tetrahedra silicate chains were also reported by Hamid \cite{Hamid1981} for tobermorite 11 \r{A} with differing calcium to silicon (Ca/Si) ratios of 0.67, 0.83 and 1.0. 

Experimental lattice parameters and atomic positions for T9\r{A} (riverside), T11\r{A} (tobermorite) and T14\r{A} (plombierite) were obtained from the American Mineralogist Crystal Structure Database \cite{Downs2003}. These materials crystallize in triclinic (space group $C\Bar{1}$), monoclinic (space group of $B11m$) and monoclinic (space group $B11b$) Bravais lattices containing $62$, $88$, and $104$ atoms per unit cell, respectively. 

The parent compound of tobermorite T11 \r{A} and T14 \r{A} shows partial occupancy for water/oxygen (WO) sites in its experimentally determined unit cell structure (i.e., the Crystallographic Information File just has WO, as it from XRD), which hinders a direct description via simulation. Refining atomic positions of the parent compound of T11\r{A} and T14\r{A}, leads to a breakdown in the nominal monoclinic symmetry to $P1$ trinclinic Bravais lattice. The angles and lattice vectors were fixed to that of the parent compound, which were used as inputs for structural optimization. The calculated structural parameters are explicitly listed in Table~\ref{tab1}. All parameters were determined using the DFT-D3 method to correct for dispersion interactions, and are compared with the available experimental data and previous DFT calculations in Table \ref{tab1}.
\begin{table}[!h]
\centering
\ra{0.5}
\caption{\label{tab1} Calculated (in italics) cell parameters of the tobermorite models. Experimentally measured values, GGA+vdW (in bold italics) and DFT-GGA values are shown for comparison. The calculated band gaps and those obtained from DFT-LDA are also listed.}
 \smallskip 
  \begin{threeparttable}
  \noindent
\begin{tabular}[t]{lllcllcll}\hline\hline
   &\multicolumn{2}{c}{Tobermorite 9 \r{A}}& \phantom{abc} &\multicolumn{2}{c}{Tobermorite 11 \r{A}}& \phantom{abc} &\multicolumn{2}{c}{Tobermorite 14 \r{A}}\\
   & \multicolumn{2}{c}{(Ca$_5$Si$_6$O$_{16}$(OH)$_2$)}& \phantom{abc} &\multicolumn{2}{c}{(Ca$_4$Si$_6$O$_{15}$(OH)$_2$.5H$_2$O)}& \phantom{abc} &\multicolumn{2}{c}{(Ca$_5$Si$_6$O$_{16}$(OH)$_2$.7H$_2$O)}\\
   & Expt. & Calc.&& Expt. & Calc.&&Expt. & Calc.\\\midrule
a(\r{A})&11.16 \tnote{a}&\it{11.19}&&6.74 \tnote{b}&\it{6.83}&&6.74\tnote{c}&\it{6.80}\\
 & & 11.21\tnote{d}&&&6.80\tnote{d}&&&6.87\tnote{d}  \\
  & & & & & 6.82\tnote{e}&&&6.64\tnote{e}\\
  & &11.27(\it{\textbf{11.22}})\tnote{f}&&&6.84(\it{\textbf{6.79}})\tnote{f}&&&\\
  \\
 b(\r{A})&7.30\tnote{a}&\it{7.32}&&7.39\tnote{b}&\it{7.48}&&7.43
 \tnote{c}&\it{7.43}\\
 &&7.39\tnote{d}&&&7.51\tnote{d}&&&7.43\tnote{d}\\
  & & & & & 7.47\tnote{e}&&&7.41\tnote{e}\\
  & &7.38(\it{\textbf{7.35}})\tnote{f}&&&7.48(\it{\textbf{7.44}})\tnote{f}&&&\\
  \\
 c(\r{A})&9.57\tnote{a}&\it{9.53}&&22.49\tnote{b}&\it{22.72}&&27.99
 \tnote{c}&\it{28.00}\\
 &&9.71\tnote{d}&&&22.57\tnote{d}&&&28.49\tnote{d}\\
  & & & & &22.70\tnote{e}&&&28.18\tnote{e}\\
  & &9.71(\it{\textbf{9.61}})\tnote{f}&&&22.72(\it{\textbf{22.60}})\tnote{f}&&&\\
  \\
 $\alpha$($^{\circ}$)&101.08\tnote{a}&\it{99.47}&&90\tnote{b}& \it{90.45}&&90\tnote{c}&\it{89.94}\\
 &&102.65\tnote{d}&&&89.83\tnote{d}&&&89.96\tnote{d}\\
  & & & & &90\tnote{e}&&&90\tnote{e}\\
  & &102.58(\it{\textbf{101.63}})\tnote{f}&&&90(\it{\textbf{90}})\tnote{f}&&&\\
  \\
 $\beta$($^{\circ}$) &92.83\tnote{a}&\it{92.27}&&90\tnote{b}&
 \it{90.01}&&90\tnote{c}&\it{90.03}\\
 &&92.54\tnote{d}&&&89.05\tnote{d}&&&90.05\tnote{d}\\
 & & & & &90\tnote{e}&&&89.99\tnote{e}\\
 & &92.01(\it{\textbf{92.23}})\tnote{f}&&&90(\it{\textbf{90}})\tnote{f}&&&\\
 \\
 $\gamma$($^{\circ}$)&89.98\tnote{a}&\it{90.60}&&123.25\tnote{b}&\it{123.13}&&123.25\tnote{c}&\it{123.93}\\
 &&89.75\tnote{d}&&&123.43\tnote{d}&&&123.47\tnote{d}\\
 & & & & & 123.22\tnote{e}&&&121.66\tnote{e}\\
 & &89.75(\it{\textbf{90.34}})\tnote{f}&&&123.57(\it{\textbf{123.63}})\tnote{f}&&&\\
 $E_{gap}$(eV)&&\it{4.52}(\it{\textbf{4.6}})\tnote{g}&&&\it{4.10}(\it{\textbf{4.2}})\tnote{g}&&&\it{4.04}(\it{\textbf{4.0}})\tnote{g}\\
 \\
 \hline\hline
    \end{tabular}
    \setlength{\tabcolsep}{1cm}
     \begin{tablenotes}
     \item[]$^a$Ref. \cite{Merlino1999}, $^b$Ref. \cite{Merlino2001}, $^c$Ref. \cite{Bonaccorsi2005}, $^d$Ref. \cite{Shahsavari2009}, $^e$Ref. \cite{Vidmer2014}, $^f$Ref.\cite{Jiang2018}, $^g$Ref.\cite{Dharmawardhana2013}
\end{tablenotes}
  \end{threeparttable}
\end{table}
The good agreement in the calculated lattice parameters relative to the available experimental values and the results cited in previous DFT calculations (deviation $\pm$1\% overall) gives confidence in the relaxed tobermorite structures. In T11\r{A} and T14\r{A}, there is a small deviation in the angles $\alpha$ and $\beta$ from the experimental measured values because of the symmetry breakdown mentioned above. 

\subsection{Electronic structure calculations}
\begin{figure}[!h]
    \centering
    (a)\includegraphics[width=0.365\textwidth]{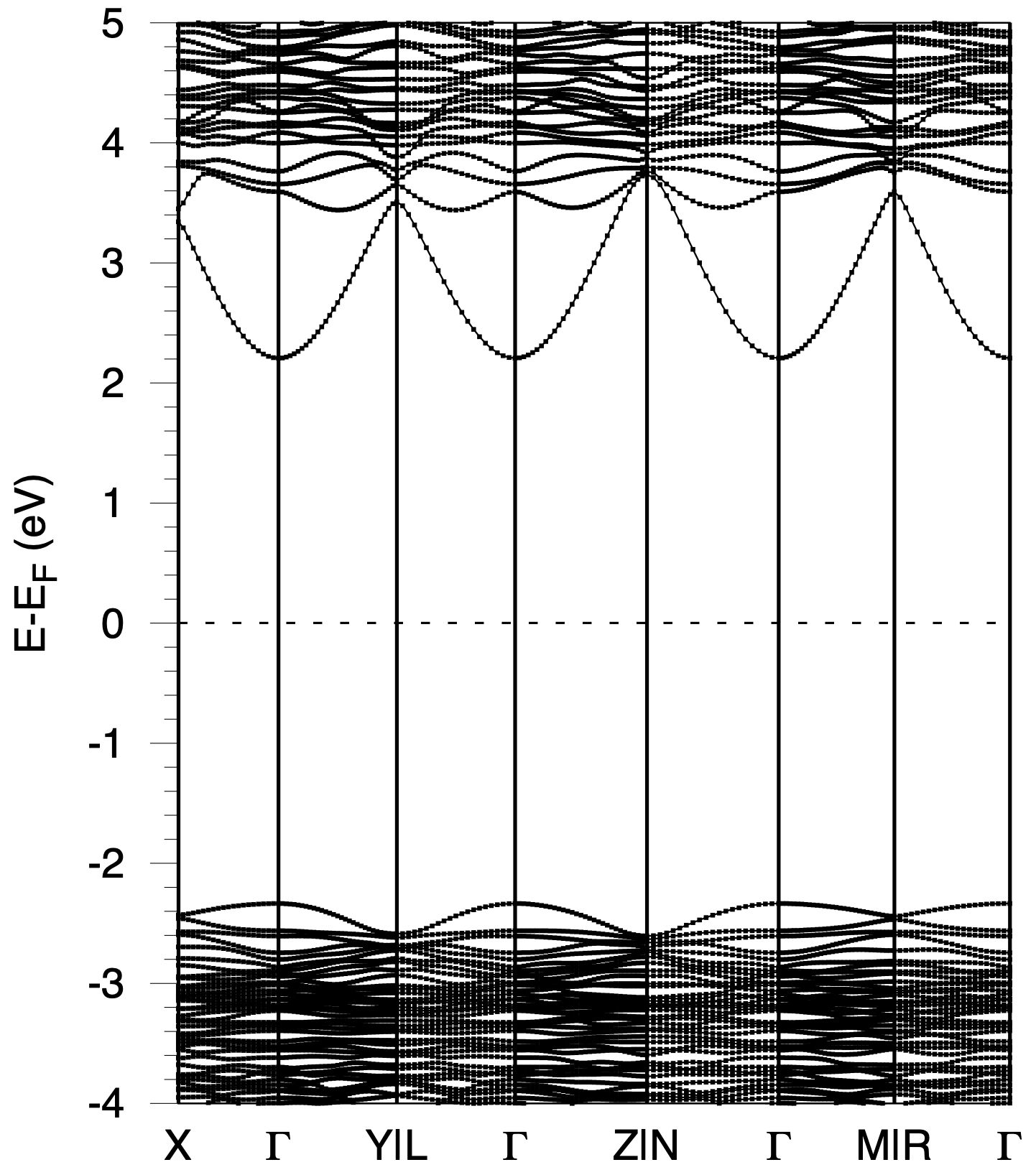} (b)\includegraphics[width=0.44\textwidth]{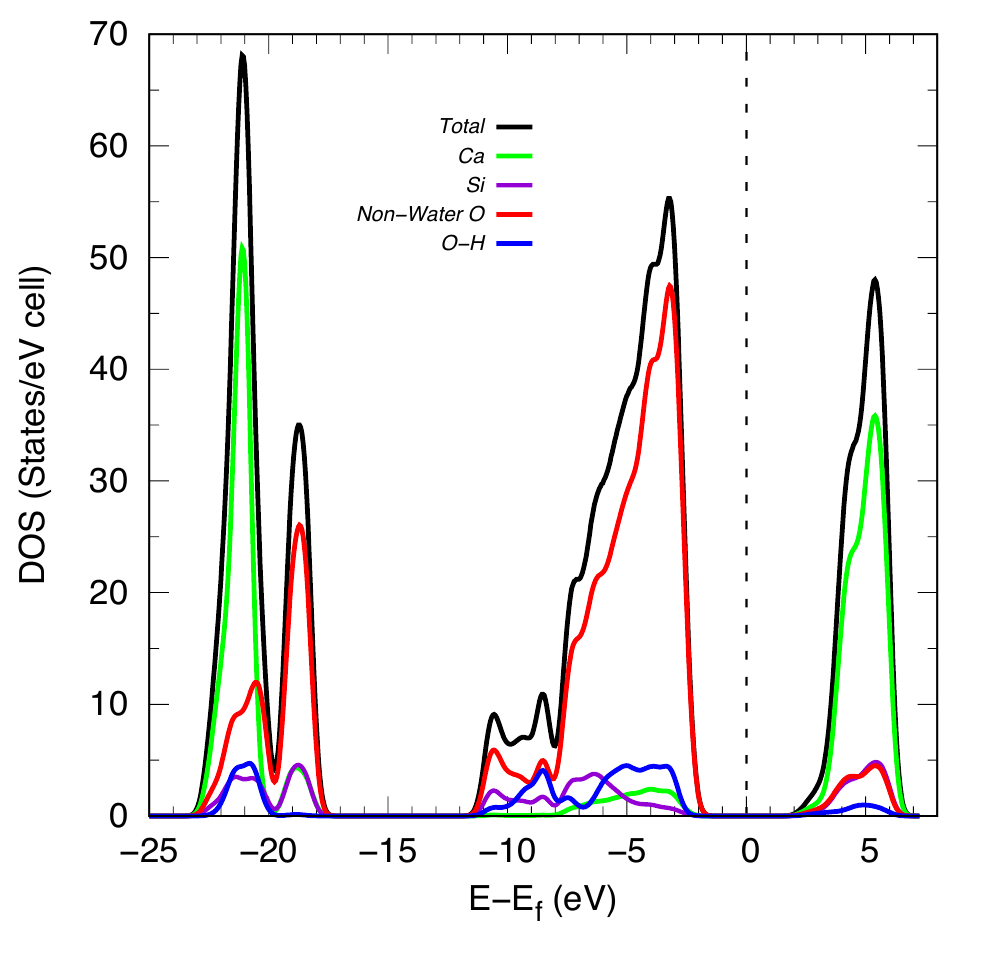}\\
    (c)\includegraphics[width=0.365\textwidth]{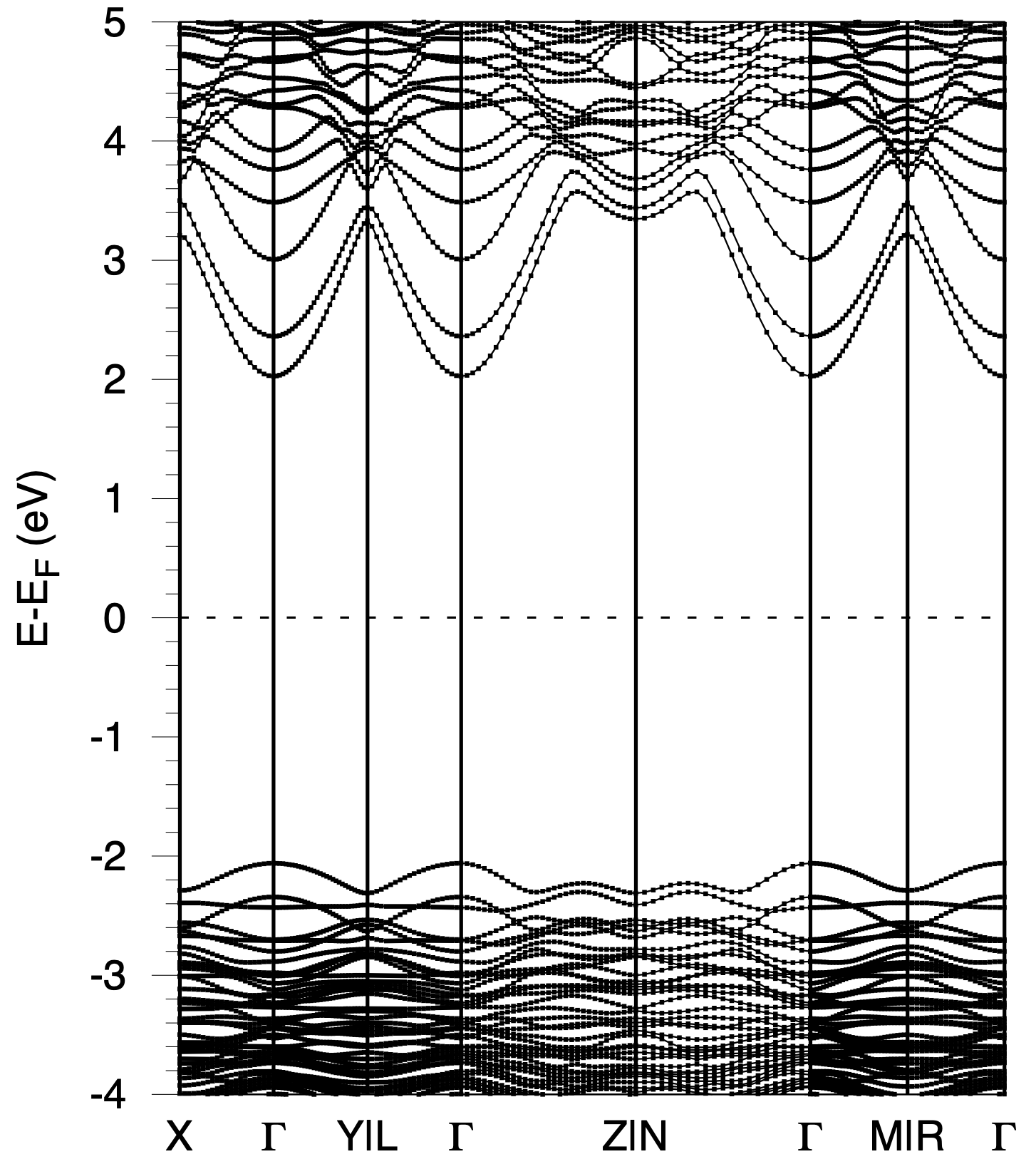} (d)\includegraphics[width=0.44\textwidth]{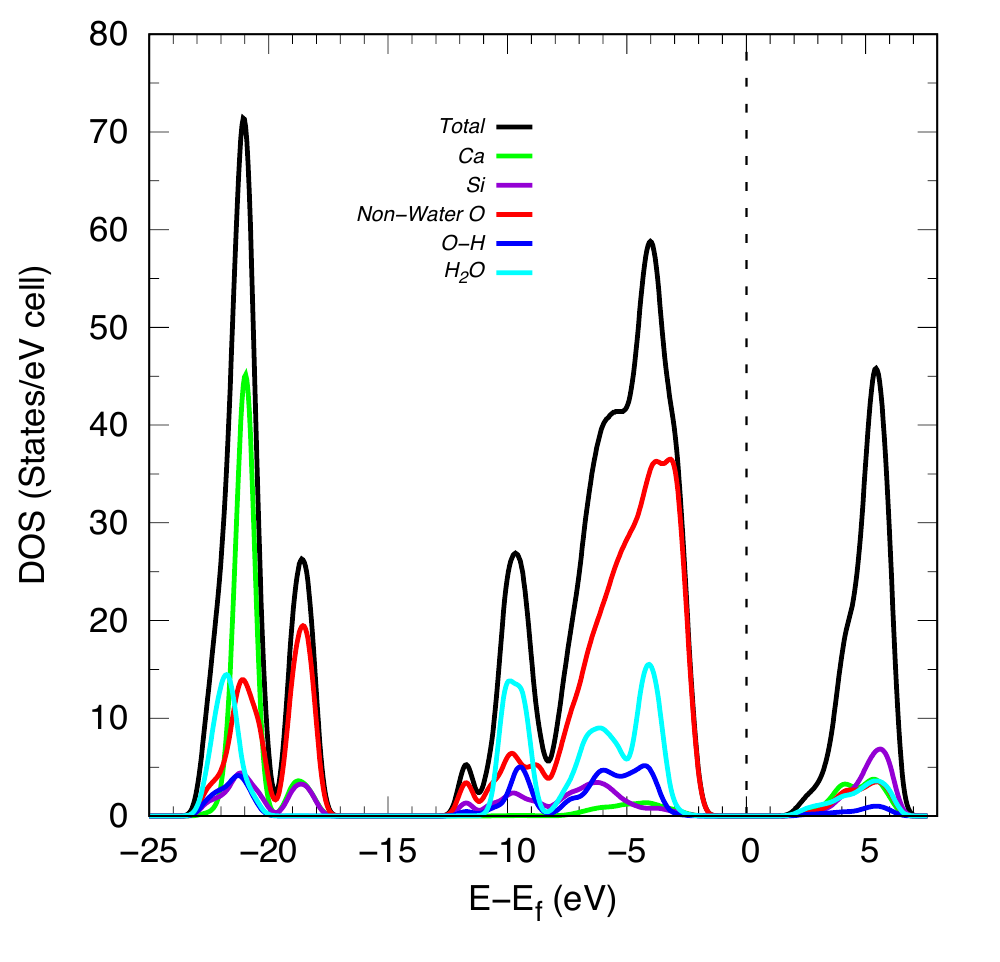}\\
    (e)\includegraphics[width=0.365\textwidth]{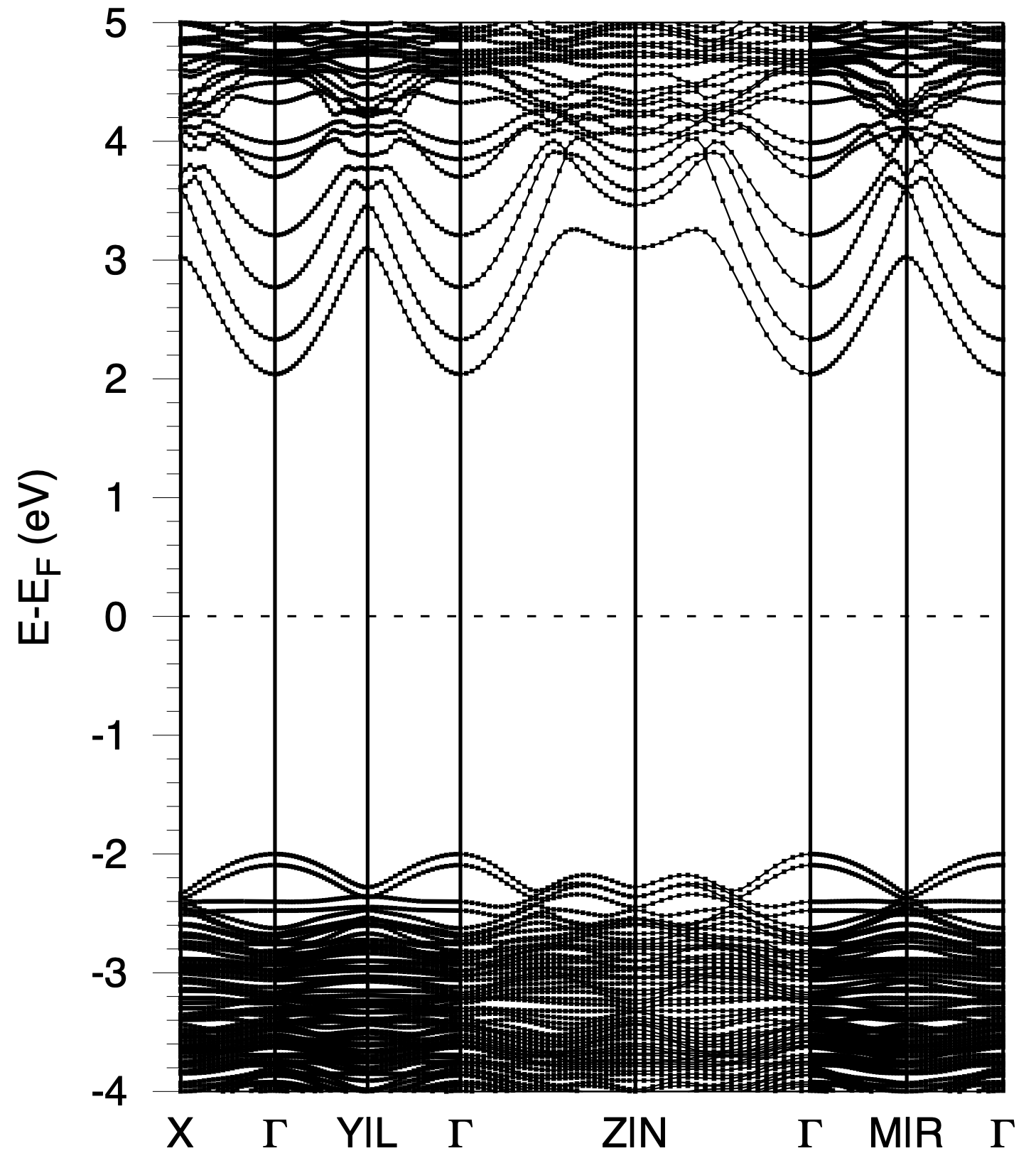} (f)\includegraphics[width=0.44\textwidth]{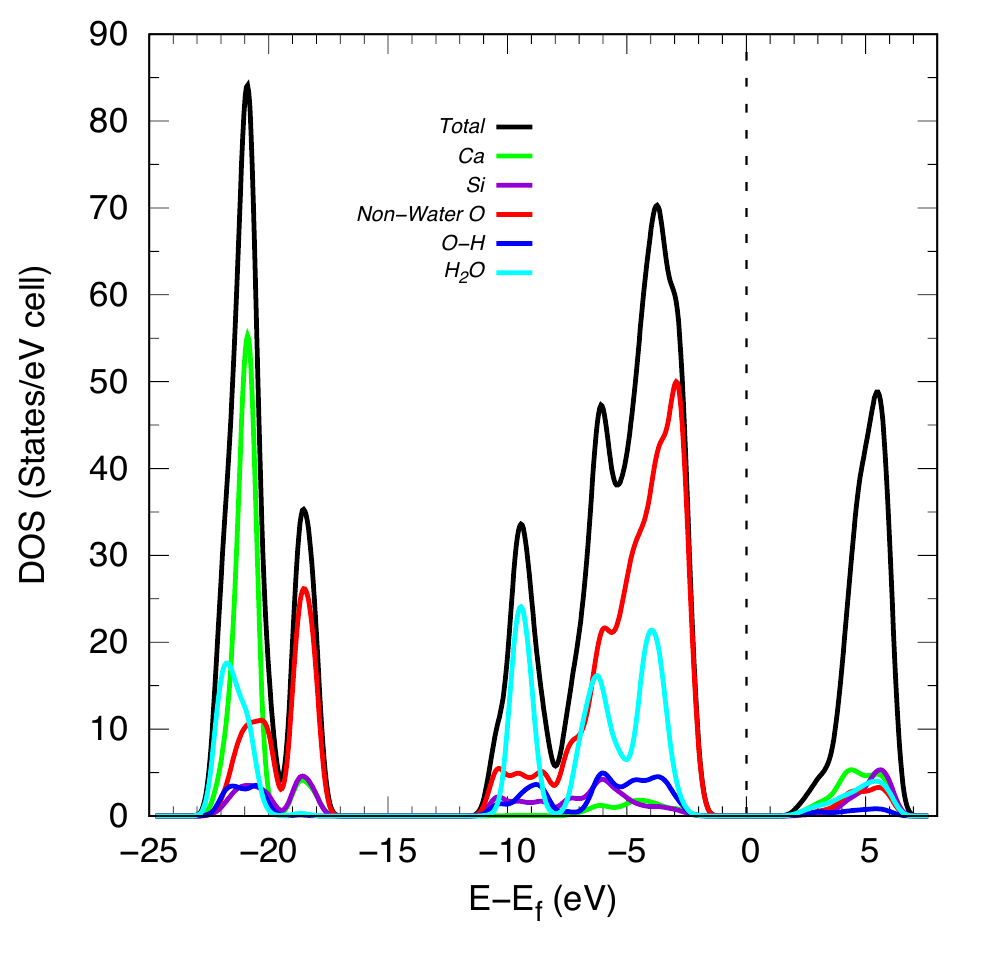}
    \caption{Calculated electronic band structures and partial density of states per element in the unit cells of T9\r{A} (a and b), T11\r{A} (c,d) and T14\r{A}(e,f). The Fermi energy level $E_f$ is set at the zero mark and lies in the middle of the band gap. The high symmetry lines for triclinic lattice were obtained from the literature \cite{Setyawan2010}.}
    \label{fig2}
\end{figure}

Accurate electronic structure calculations were carried out to obtain the transport properties.
These calculations were performed using the relaxed cell parameters listed in Table~\ref{tab1}. The calculated band structures of the tobermorite models along some high symmetry paths for triclinic lattices \cite{Setyawan2010} are presented in Fig.~\ref{fig2}. 

The tobermorite models are insulators and generally expected to have wide energy band gaps. The band structures of all the tobermorite models showed a direct energy band gap at the $\Gamma$-point. The resulting gaps ($E_{gap}$) listed in the last row of Table~\ref{tab1}, i.e. 4.6 eV, 4.2 eV and 4.0 eV for T9\r{A}, T11\r{A} and T14\r{A}, respectively, were found to be in very good agreement with previous DFT-LDA results \cite{Dharmawardhana2013}.  

Our band structure calculations showed that as the water content increases in T11\r{A} and T14\r{A} structures, their band gaps decrease as compared to T9\r{A}, which contains no water molecules. Most of the bands near the valence band maximum (VBM) are flatter (heavy bands) than the bands near the conduction band minimum (CBM), which indicate larger effective mass for the holes than for the electrons. The resulting heavy bands near the VBM should lead to a large Seebeck coefficient in $p$-type thermoelectric materials. It is interesting to note that the conduction bands (Fig.~\ref{fig2}) of the tobermorite models exhibit obvious parabolic character, which suggests that their transport properties can also be modeled using the simple parabolic band approach.

To gain a better understanding of the nature of the states in the VBM and CBM, we have plotted the element-resolved partial density of states (pDOS) of the unit cells of T9\r{A}, T11\r{A} and T14\r{A}, respectively, as shown in Fig.~\ref{fig2}. The sharp peaks in the electronic density of states below the Fermi level for all three compounds originate almost entirely from the non-water O states, with small contributions from the water O-H bonds, Si, Ca and H$_2$O states, respectively. From Fig.~\ref{fig2}, it can be seen that non-water oxygen O-$p$ states contribute significantly to the pDOS between -8.5 eV to -2.0 eV, hybridizing with the Si-$p$ states and forming strong covalent bonds that lead to the formation of SiO$_4$ tetrahedra. The Ca-$p$ states exhibit a relatively small contribution to the density of states (from -7 eV to -3 eV), which indicates a negligible contribution of the alkaline-earth metal to the covalent bond but stresses the ionic bonding within the interlayer calcium polyhedra. This ionic bond character is greater in T9\r{A} than in T11\r{A} and T14\r{A} models due to the absence of H$_2$O molecules. The hybridized pDOS is expected to be beneficial to the electronic transport properties. 
In the tobermorite models, the water molecules are confined to the interlayer region and interact in the sense of hybridising with all other elements within the crystal. The peak in the lower valence band composed mainly of the Ca-$s$ states was observed to make significant contributions from -24 eV to -20 eV and their interaction with the non-water O contributes to the ionic bonding within the interlayers of Ca-O polyhedra. Turning to the conduction bands, the Ca and Si states dominate the conduction band minimum. These bands were observed to be generally more dispersive than those found in the valence bands. Dispersive bands are considered to contribute to high electron mobility because of the small effective mass of the carriers.
\subsection{Electronic transport properties}
\subsubsection*{Seebeck Coefficient}
\begin{figure}[t]
    \centering
    \includegraphics[width=0.6\textwidth]{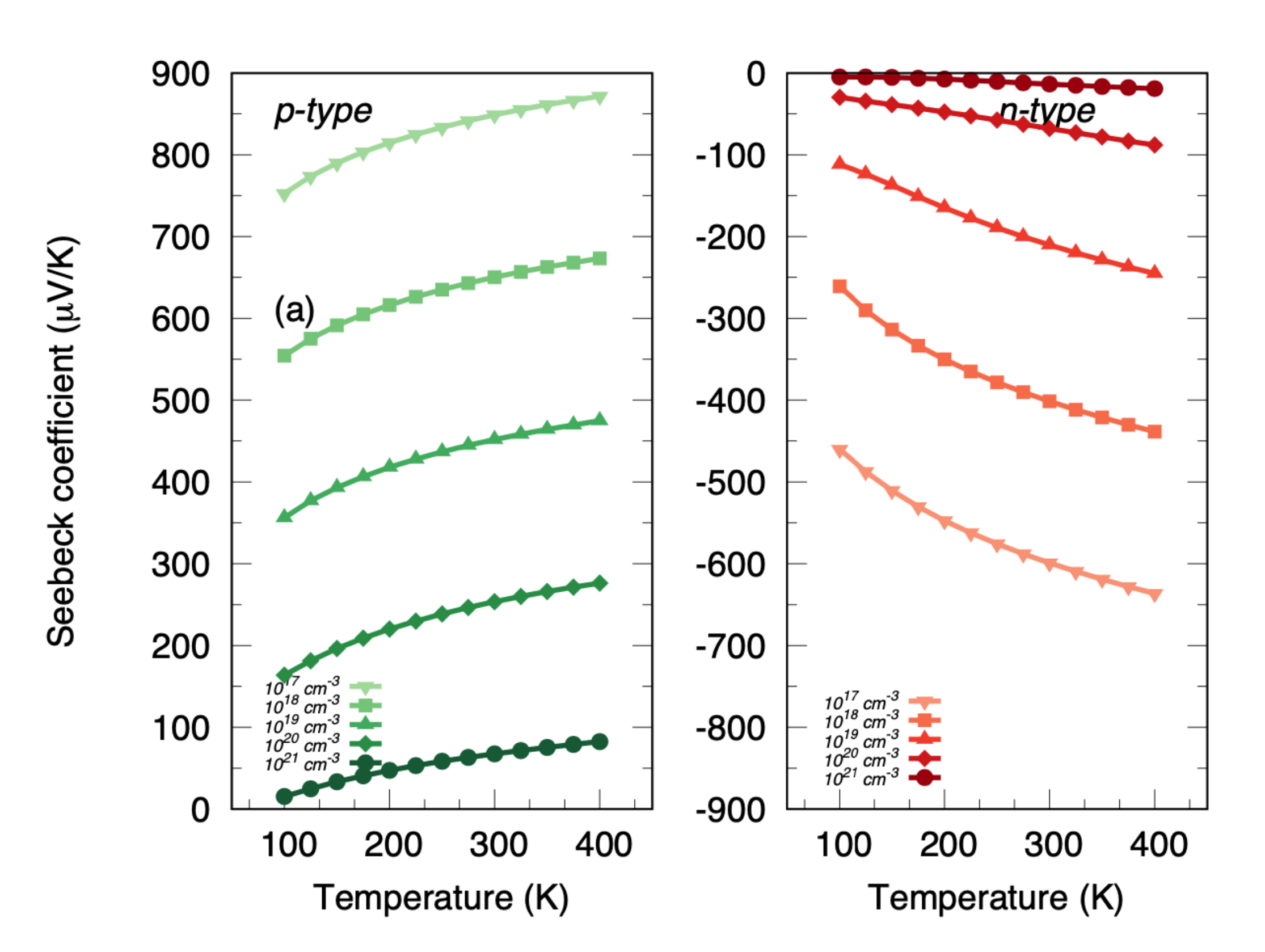}\\
    \includegraphics[width=0.6\textwidth]{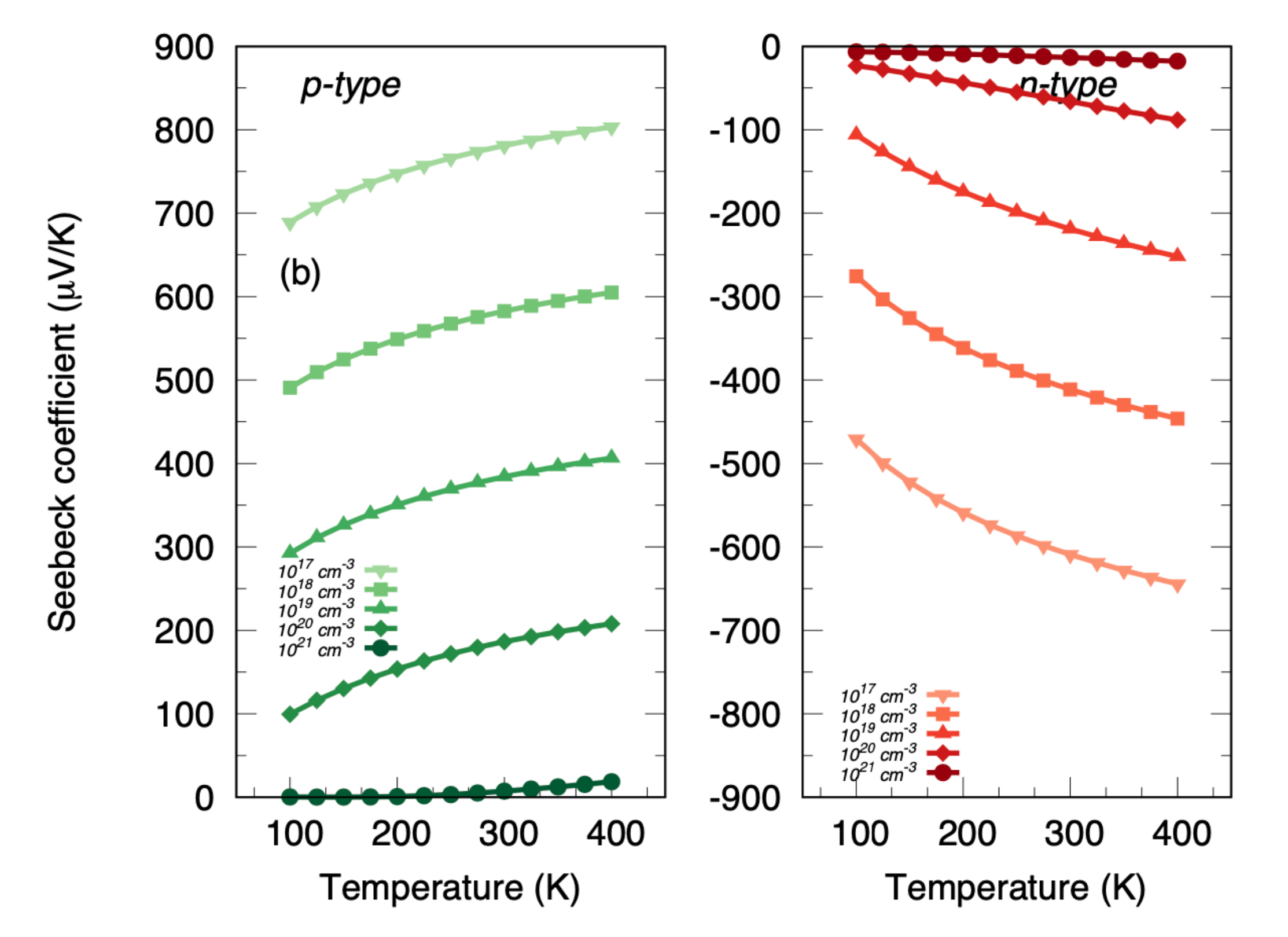} 
    \includegraphics[width=0.6\textwidth]{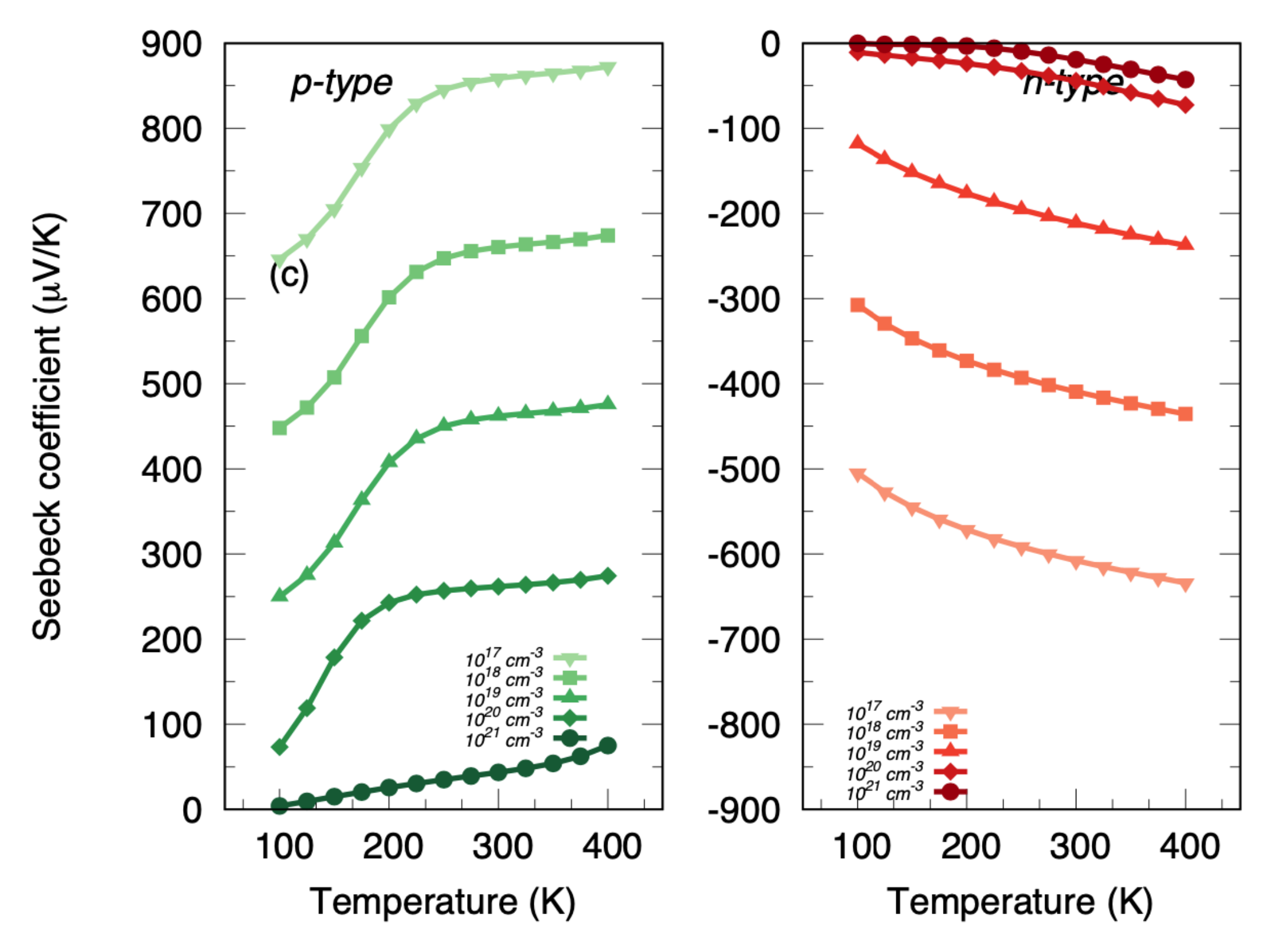}
    \caption{Calculated averaged Seebeck coefficient with respect to temperature at fixed carrier concentrations in $p$-type and $n$-type, for T9\r{A}(a), T11\r{A}(b) and T14\r{A}(c). }
    \label{fig4}
\end{figure}

Within the constant relaxation time approximation (CRTA), the Seebeck coefficient $S$ is independent of $\tau$. Without any adjustable parameters, $S$ results exclusively from a first principles electronic structure calculation \cite{Snyder2008}. For  a parabolic band model of a wide band-gap or non-degenerate semiconductor, the Seebeck coefficient  is given by \cite{May2009,Bhattacharya2011}:
\begin{eqnarray}
\label{eqn8}
S &=&\pm \frac{k_{B}}{q}\left[\eta-\left(r + \frac{5}{2}\right) \right],
\end{eqnarray}
where $\eta$ is the reduced carrier energy, $r$ is the scattering parameter, $q$ is the electron charge and $k_B$ is the Boltzmann constant. Fig.~\ref{fig4} shows the temperature dependence of the averaged Seebeck coefficient of all tobermorite models for different values of carrier concentrations ranging from $10^{17}$ to $10^{21}$ cm$^{-3}$. These calculations were performed for both the $p$-type and $n$-type doping. The averaged Seebeck coefficient is a conductivity-weighted average of its directional components \cite{Parker2013}:
\begin{eqnarray}
S_{avg} = \dfrac{S_a\sigma_a + S_b\sigma_b + S_c\sigma_c}{\sigma_a+\sigma_b+\sigma_c},
\end{eqnarray}
which stem from the definition of $S$ in terms of the Onsager coefficients: $S=L_1/L_0T$. Here, $a$, $b$, and $c$ are the Cartesian axes. This equation is the direct expression of the directional average of the transport coefficient in Eq.~(\ref{eqn5}), provided that the relaxation time is independent of the direction. It is important to note that the negative sign of the Seebeck coefficient indicates electron carriers while the positive sign is for hole carriers. For the sake of simplicity, our discussion will be focused on the values rather than the signs. The calculated values of Seebeck coefficient are almost invariant in both $p$-type and $n$-type doping and grow monotonically with temperature as the carrier concentration increases from $10^{17}$ to $10^{21}$ cm$^{-3}$ for T9\r{A}, T11\r{A} and T14\r{A}, respectively. In all the tobermorite models, the calculated Seebeck coefficients show a linear decrease in their absolute values with increasing carrier concentrations as required by Eq.~(\ref{eqn8}). The optimal doping for thermoelectric performance for these models expected at $10^{17}$ cm$^{-3}$ for both $p$-type and $n$-type materials. Furthermore, from Fig.~\ref{fig4} the calculated absolute Seebeck coefficients for $p$-type doping in the tobermorite models were observed to be larger than those of the $n$-type doping. This is the direct consequence of the heavy valence bands present in the band structure of the tobermorite models in which the effective mass of the hole carriers keeps increasing. T14\r{A} (Fig.~\ref{fig4}(c)) shows higher $S$ values at lower carrier concentration for $p$-type doping than those of T9\r{A} (Fig.~\ref{fig4}(a)) and T11\r{A} (Fig.~\ref{fig4}(b)), because of its smaller energy band gap compared to the other two models.
These values of $S$ for tobermorite models are considered the first results, since there are no available experimental or theoretical data to compare them with.
\subsubsection*{Relaxation Time}
\label{Sec:RelaxTime}
The BoltzTraP code\cite{Madsen2006} returns the electric conductivity divided by the constant relaxation time $\sigma/\tau$. To obtain a quantitative estimate of the electrical conductivity $\sigma$, we need to calculate the relaxation time $\tau$ independently. Considering the high computational cost and the size of the unit cells of the tobermorite models, only the energy-dependent relaxation time for the T9\r{A} model ---\emph{i.e.}, the model with the fewest atoms in the unit cell --- was calculated using the simple parabolic band (SPB) approach \cite{Nag1980,Lee2016}, as a function of temperature, and for carrier concentrations ranging from $10^{17}$ to $10^{21}$ cm$^{-3}$ for both electrons and holes.

In the SPB approximation, the relaxation time is assumed to follow a power-law expression:
$\tau=\tau_{0}\left(T\right) \left(\epsilon-\epsilon_0\right)^{r-1/2}$, where $\epsilon_0$ is a reference energy, \emph{e.g.} the valence band maximum (VBM) or the conduction band minimum (CBM). 
The value $r=0$ is appropriate for scattering from acoustic phonons at a temperature larger than the Debye temperature. Note that the prefactor $\tau_{0}$ is temperature dependent (see below).
The SPB approach has been successfully applied to study the thermoelectric properties of materials as reported in the literature \cite{May2009,Bottger2011,Xi2012,Liu2014,Guo2015}. 

In the case of (implicit) $n$-type doping, the Fermi energy $E_F$ was first determined from the carrier concentration given by
\begin{eqnarray}
n\left(T\right) &=& \frac{1}{2\pi^2}\left(\frac{2m^*_dk_BT}{\hbar^2}\right)^{\frac{3}{2}}F_{\frac{1}{2}}(\eta),
\end{eqnarray}
and  the Fermi integral $F_j(\eta)$ is defined as
\begin{eqnarray}
\label{eqn4}
F_j(\eta) &=& \int_0^\infty \frac{\epsilon^j}{1+\exp{\left(\frac{\epsilon}{k_BT}-\eta\right)}}d\epsilon\;, 
\end{eqnarray}
where $\eta=(E_F-\epsilon_0)/k_BT$. For electrons, $\epsilon_0$ is the energy of the CBM. The density of states effective mass is defined as $m^*_d = N_v^{\frac{2}{3}}(m^*_a m^*_b m^*_c)^{\frac{1}{3}}$, where $N_v$ is the degeneracy of the band valley and $m^*_a$, $m^*_b$, $m^*_c$ are the principal effective masses with respect to the Cartesian directions at the band valley. The principal effective masses were evaluated from the Hessian matrix of band energies at the band edges using:
\begin{eqnarray}
\frac{1}{m^*_{ij}}=\frac{1}{\hbar^2}\frac{\partial^2\epsilon}{\partial k_i\partial k_j}
\end{eqnarray}
where $i,j$ are Cartesian components. The elements of the Hessian matrix were calculated using the finite difference method with a $5$-point central difference stencil \cite{Fonari2012}. 
The calculated effective masses at the conduction band minimum are $m^*_{k_a}=0.149 m_0$, $m^*_{k_b}=133 m_0$ and $m^*_{k_c}=0.155 m_0$, where $m_0=9.11\cdot10^{-31}$ kg is the free electron mass.
Similar calculations were done in the case of (implicit) $p$-type doping. In this case the calculated effective masses at the valence band maximum are $m^*_{k_a}=0.309m_0$, $m^*_{k_b}=1.161m_0$ and $m^*_{k_c}=0.989m_0$.

According to the SPB approach, the energy dependent relaxation time for phonon scattering is expressed as:\cite{Guo2015}
\begin{eqnarray}\label{tau_acoustic}
\tau &=& \frac{\sqrt{2}\pi \hbar^4 v_s^2\rho}{3E_d^2\left(m^*k_BT\right)^{\frac{3}{2}}}\frac{F_0(\eta)}{F_{\frac{1}{2}}(\eta)}, 
\end{eqnarray}
where we consider acoustic phonons because their scattering is dominant at temperatures larger than the Debye temperature and moderate doping\cite{Lee2016}.
Eq.~(\ref{tau_acoustic}) is consistent with the SPB approximation if we take $1/\tau_0\propto k_BT$, i.e., the scattering rate proportional to the number of phonons.
Here, $\rho$ is the mass density and the longitudinal sound velocity $v_s$ is calculated from the elastic constants values of $(C_{11}/\rho)^{\frac{1}{2}}$, $(C_{22}/\rho)^{\frac{1}{2}}$ and $(C_{33}/\rho)^{\frac{1}{2}}$ in the direction of the wave vector \cite{Anderson1963}. The elastic constants ($C_{11}$=158 GPa, $C_{22}$=154 GPa and $C_{33}$=74 GPa) were obtained from a DFT calculation at the gamma point. $E_d$ is the acoustic deformation energy defined by $E^i_d = \Delta\epsilon_i/(\Delta V/V_0)$ and calculated from the energy change ($\Delta\epsilon_i$) of the $i$-$th$ band with volume dilation ($\Delta V/V_0$) along the Cartesian directions. We calculated the band structure by varying the volume from $0.98V_0$ to $1.01V_0$ with $\pm$5.0\% increase of the cell volume. To calculate the deformation energy, we adopted the method proposed by Xi $et$ $al.$ \cite{Xi2012} by taking the energy change of the conduction band minimum (CBM) and the valence band maximum (VBM) for electrons and holes, respectively.
The evaluated deformation energy for T9\r{A} is 10.428 eV for the VBM and 5.381 eV for the CBM. 

\begin{figure}[!h]
    \centering
    (i)\includegraphics[width=0.4\textwidth]{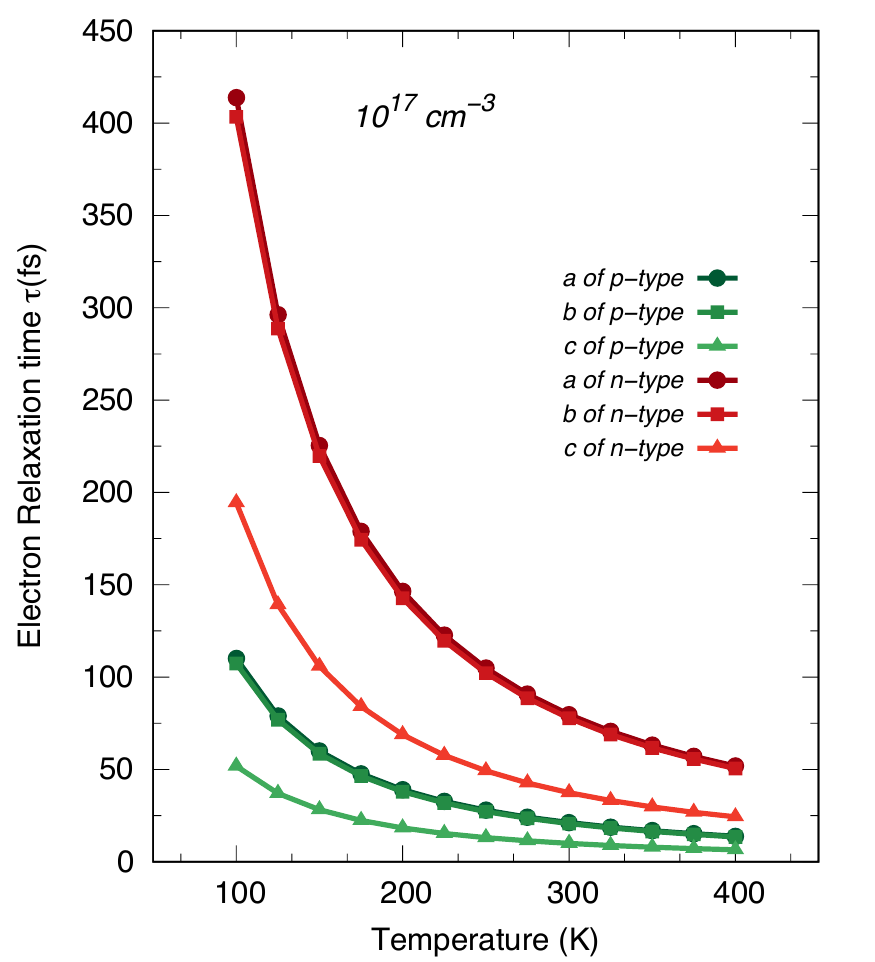} (ii)\includegraphics[width=0.4\textwidth]{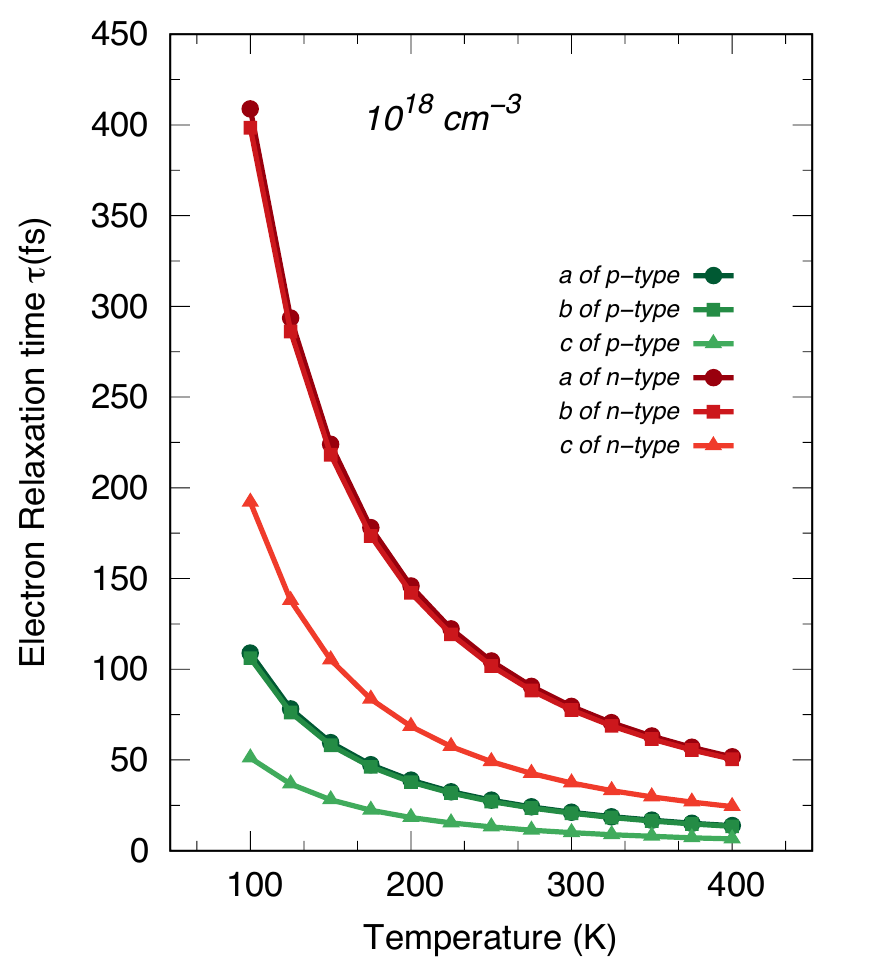}\\
    (iii)\includegraphics[width=0.4\textwidth]{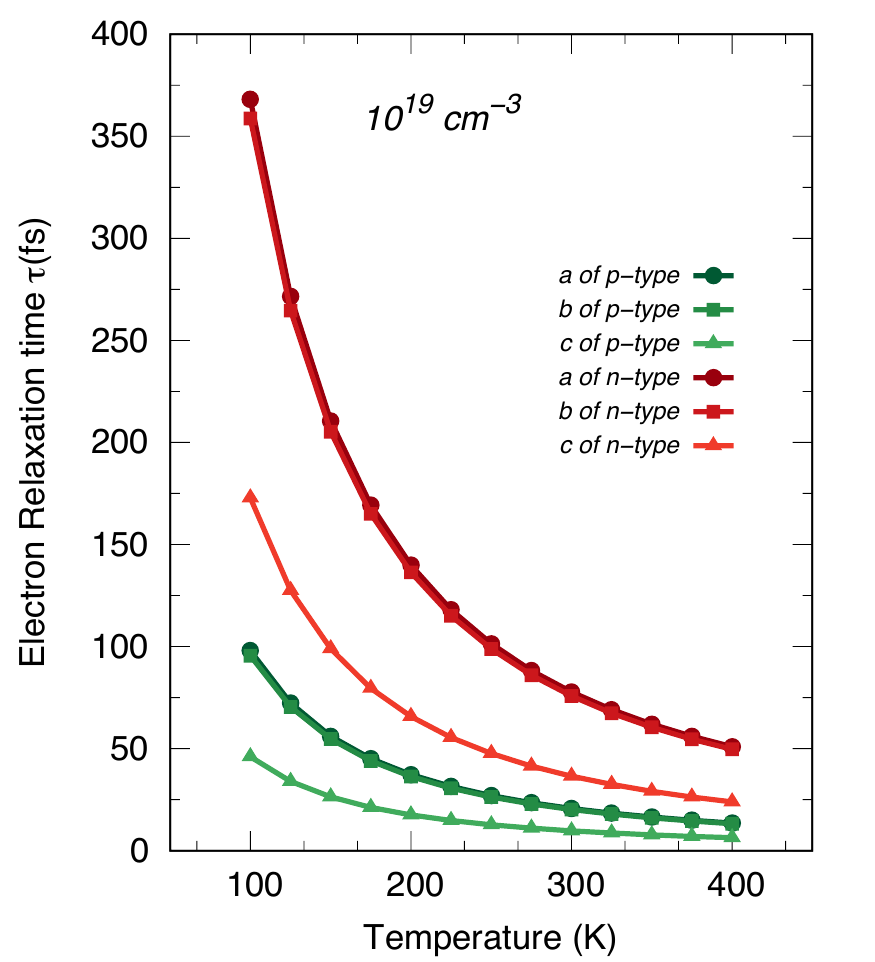} (iv)\includegraphics[width=0.4\textwidth]{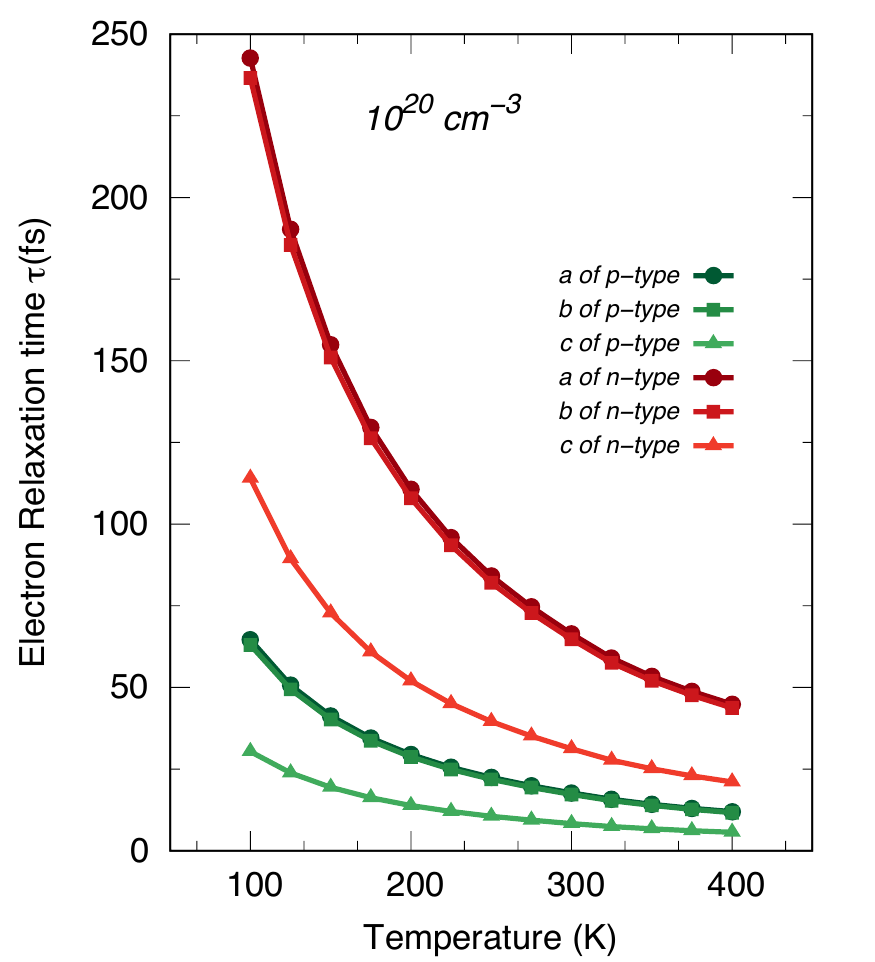}\\
    (v)\includegraphics[width=0.4\textwidth]{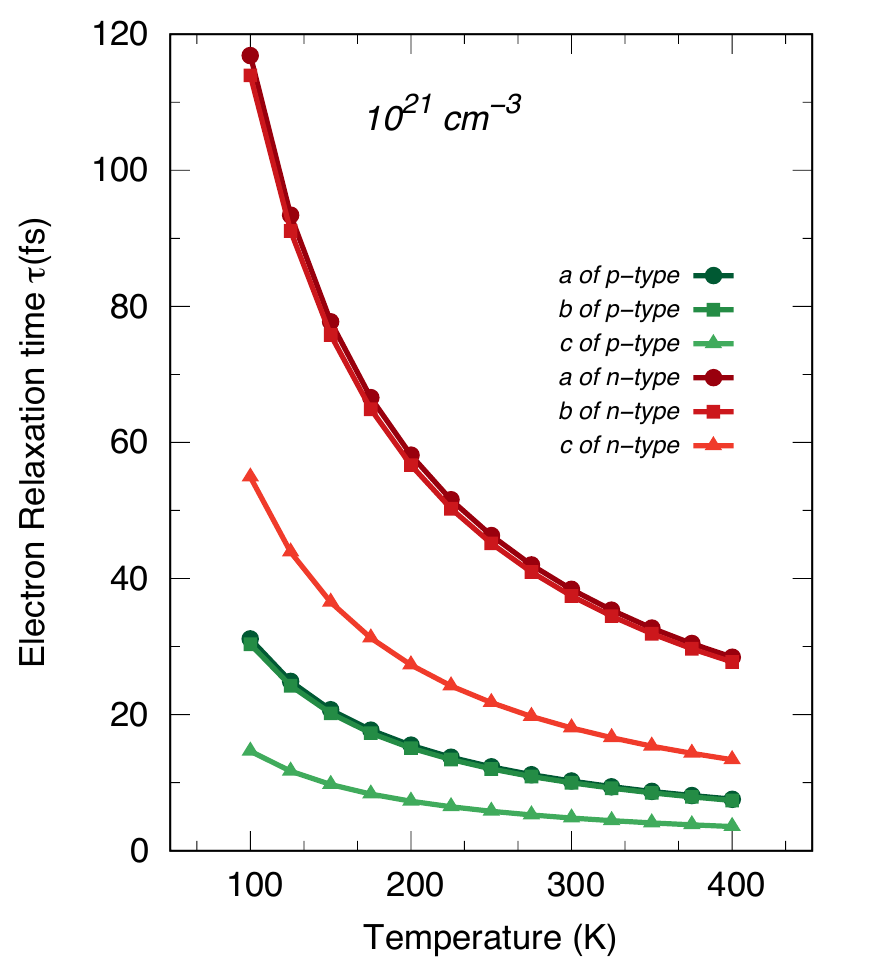} \\
    \caption{Calculated electronic relaxation time $\tau$ versus temperature for T9\r{A}.}
    \label{fig5}
\end{figure}

In Figs.~\ref{fig5}(i) to \ref{fig5}(v), we plot the relaxation time $\tau$ as a function of temperature and carrier concentration obtained from the SPB model for T9\r{A}, along the three Cartesian axes. A glance at Fig.~\ref{fig5} shows that at low temperatures $\tau$ is large and decreases with increasing temperature. As we move from $10^{17}$ cm$^{-3}$ to higher carrier concentrations, the relaxation time $\tau$ decreases linearly with temperature for both $p$-type and $n$-type doping. The $\tau$ values along the $x$ and $y$ directions are comparable because the longitudinal sound velocity is nearly the same. Over the entire carrier concentration range, the $n$-type relaxation time $\tau$ exhibits higher values compared to those of the $p$-type. This is because the electrons have a much smaller effective mass than the holes, which results in a lower density of states and hence a longer relaxation time $\tau$.

\subsubsection*{Electrical Conductivity} 
\begin{figure}
    \centering
    (i)\includegraphics[width=0.6\textwidth]{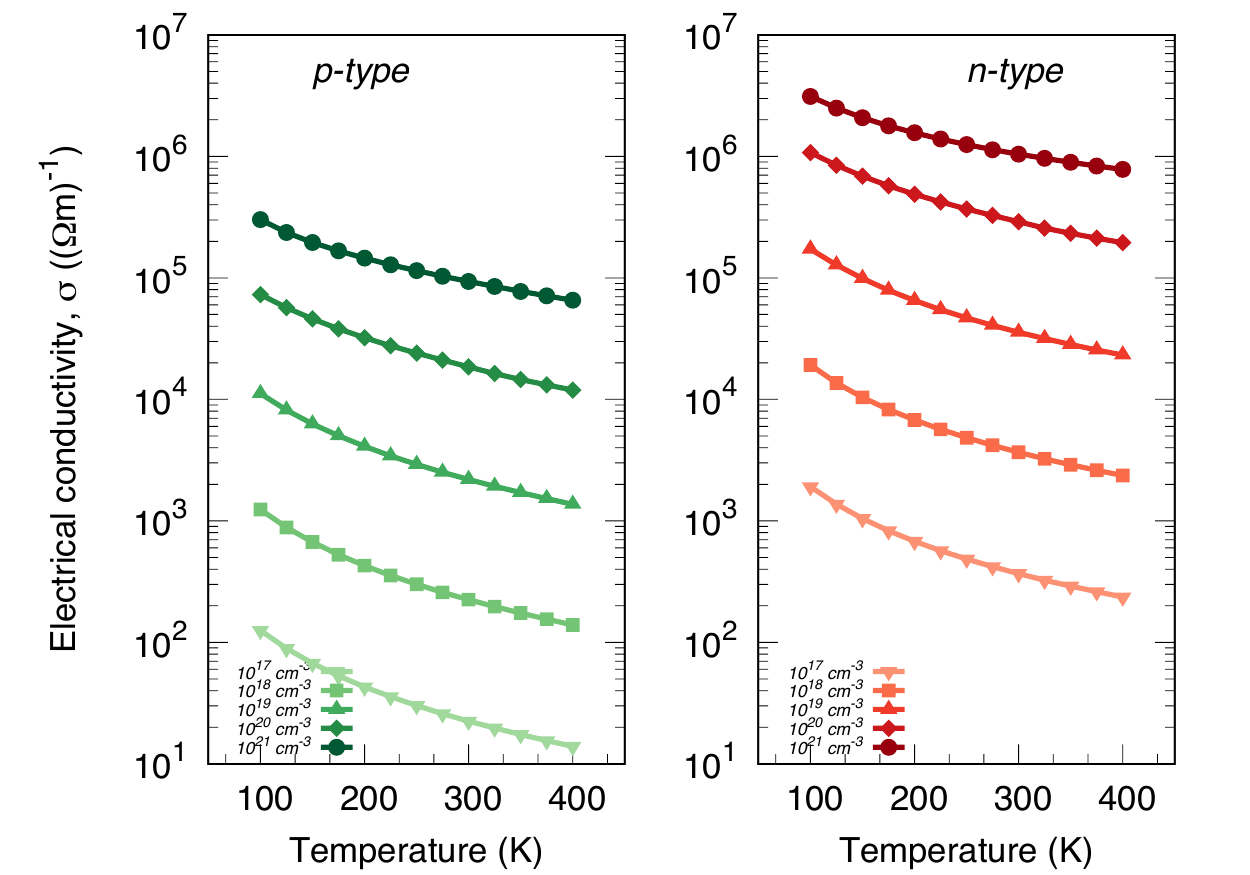}\\ 
    (ii)\includegraphics[width=0.6\textwidth]{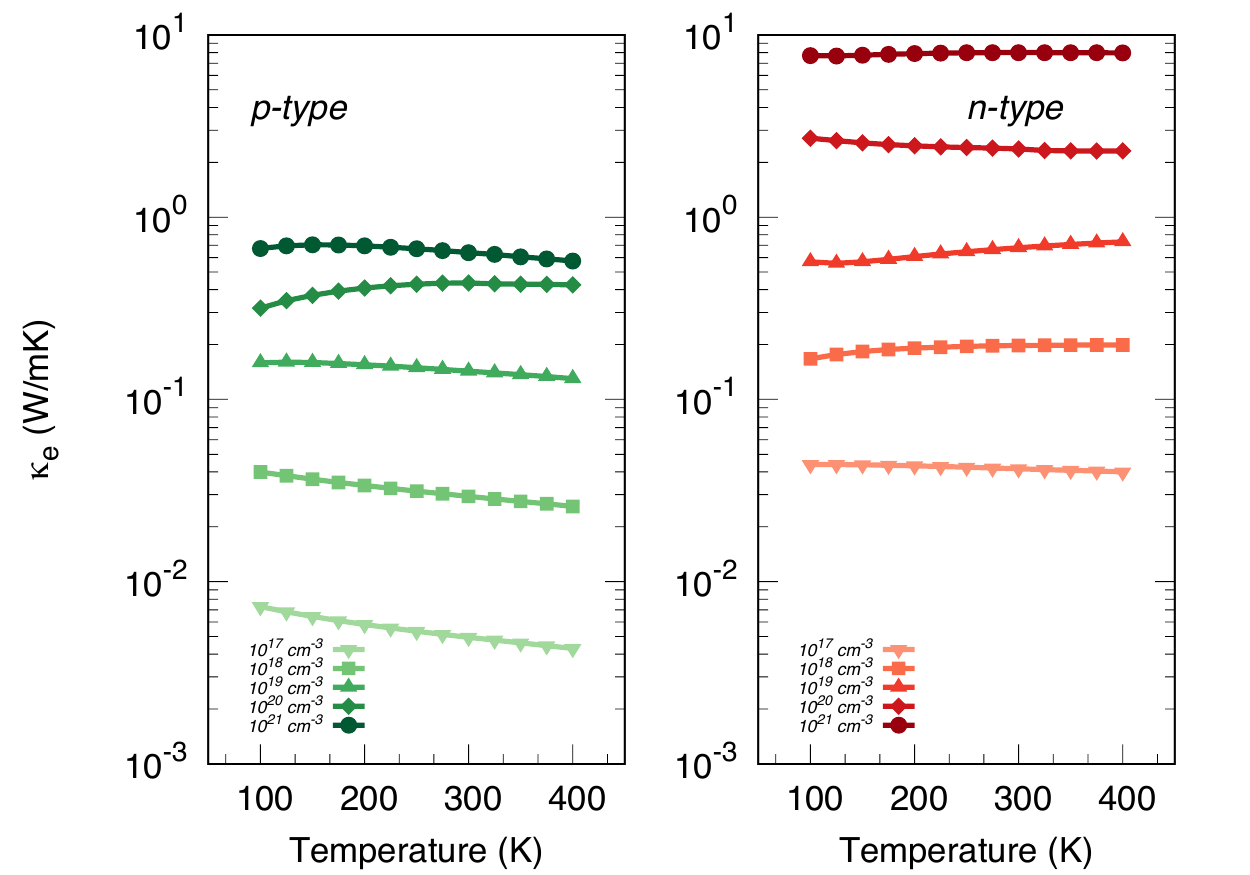}\\
    \caption{Calculated averaged electrical conductivity and electronic thermal conductivity as a function of temperature and carrier concentrations for T9\r{A}.}
\label{fig6}
\end{figure}

\begin{figure}[!h]
    \centering
    (i)\includegraphics[width=0.655\textwidth]{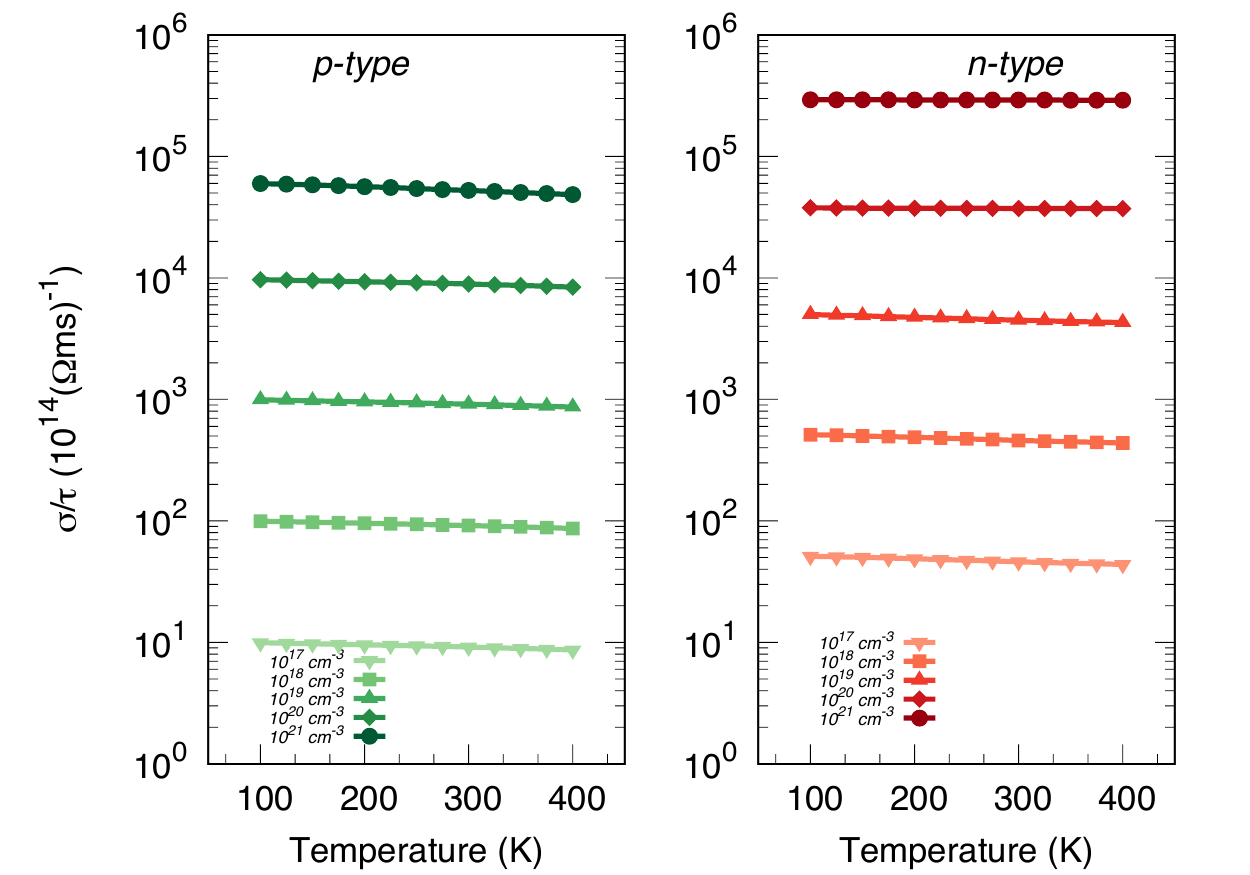}\\ 
    (ii)\includegraphics[width=0.655\textwidth]{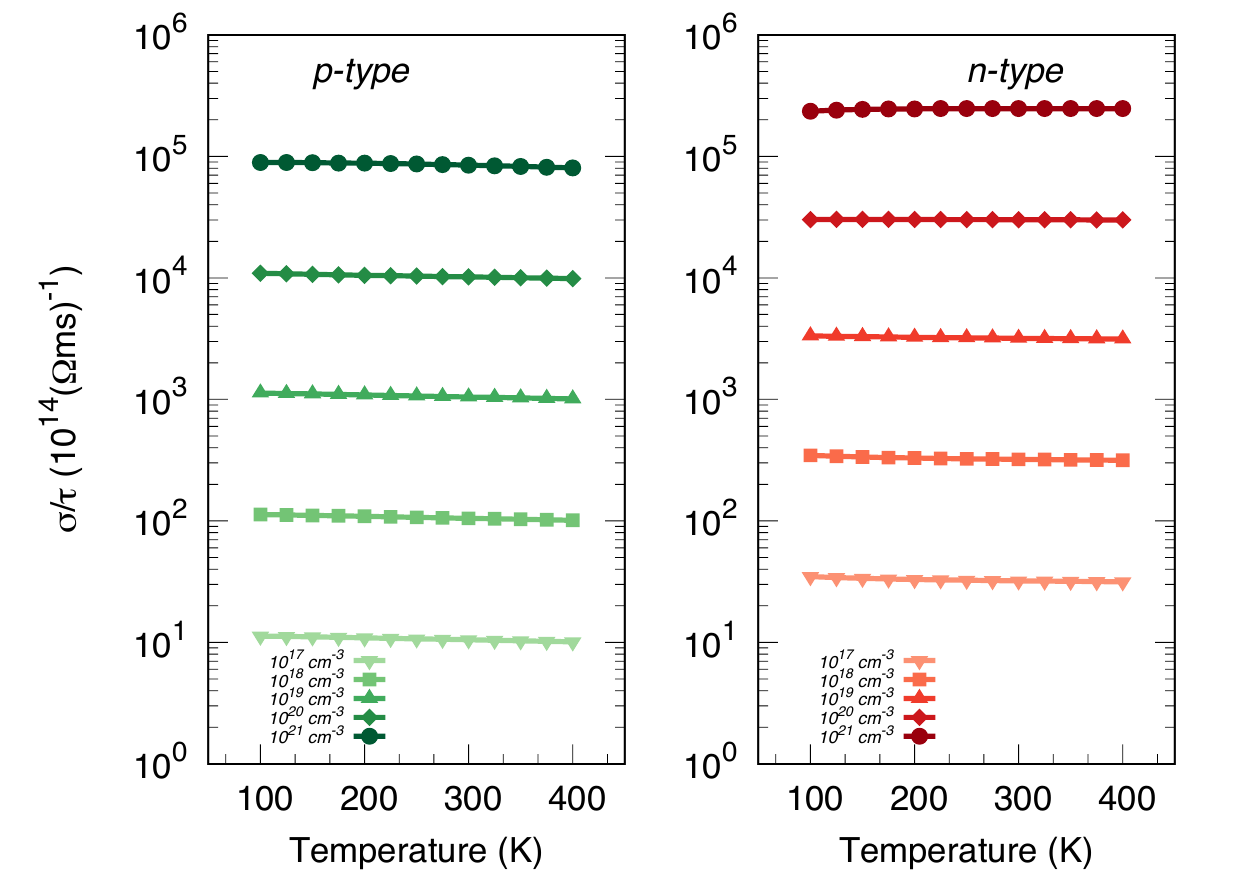}\\
    \caption{Averaged electrical conductivity $\sigma/\tau$, as a function of temperature and carrier concentrations for T11\r{A}(i) and T14\r{A}(ii) models.}
    \label{fig7}
\end{figure}

The trace of the electrical conductivity ${\sigma}_{avg}$=$\left(\sigma_a+\sigma_b+\sigma_c\right)/3$ was obtained from BoltzTraP for T9\r{A} using the calculated relaxation time computed in Sec.~\ref{Sec:RelaxTime}. The resulting electrical conductivity $\sigma$ for $p$-type and $n$-type doping as a function of temperature and fixed carrier concentrations ($10^{17}$ to $10^{21}$ cm$^{-3}$) is presented in Fig.~\ref{fig6}. Both $p$-type and $n$-type T9\r{A} exhibit an increase in electrical conductivity with increasing carrier concentration, which is consistent with the dependence of electrical conductivity on the carrier concentration. At $300$ K, we found $p$-type electrical conductivity values of $22.49$, $224.43$, $2190.98$, $18534.20$ and $93440.60$ $\Omega^{-1}$m$^{-1}$ for carrier concentrations increasing from $10^{17}$ cm$^{-3}$ to $10^{21}$ cm$^{-3}$. For $n$-type doping, electrical conductivity values of $367.62$, $3666.00$, $30631.10$, $289992.00$ and $1.044\times10^6$ $\Omega^{-1}$m$^{-1}$ were obtained for the same concentration range at $300$ K. A further glance at Fig.~\ref{fig6} shows that at low temperature $\sigma$ is higher and then decreases as temperature increases, a behaviour attributed to the temperature dependence of the electron-phonon scattering described by a T$^{-3/2}$ power-law. However, the electrical conductivity increases overall with increasing carrier concentration from $10^{17}$ to $10^{21}$ cm$^{-3}$ in both the $p$-type and $n$-type, respectively. This is typical of a lightly doped semiconductor where the electrical conductivity decreases with temperature when most of the carriers are ionised, as a result of the thermal scattering of carriers by the vibrating lattice \cite{Kasap2005}. The overall $n$-type electrical conductivity $\sigma$ values are significantly higher than those of the $p$-type doping at the same carrier concentration and temperature. This is due to the smaller effective mass at the conduction band maximum of the $n$-type doping. 
\\
Fig.~\ref{fig7} shows the temperature-dependent electrical conductivity $\sigma/\tau$ versus carrier concentration obtained from BoltzTraP, in both $p$-type and $n$-type doping for T11\r{A} and T14\r{A} models.
The general observation from Figs.~\ref{fig7}(i) and \ref{fig7}(ii), is that $\sigma/\tau$ decreases at a negligible rate with temperature over the entire carrier concentrations range, reaching a maximum value of $<0.1\times10^{6}$ $(10^{14}(\Omega ms)^{-1})$ for $p$-type and $>0.2\times10^{6}$ $(10^{14}(\Omega ms)^{-1})$ for $n$-type at carrier concentrations of $10^{21}$ cm$^{-3}$, for T11\r{A} and T14\r{A} models, respectively. In the whole carrier concentration range, the $n$-type values are higher than those of $p$-type doping in both T11\r{A} and T14\r{A}. This difference can be attributed to the heavy valence band maximum in their band structures (see Fig.~\ref{fig2}(b) and \ref{fig2}(c)), which causes a reduction in the mobility of the hole carriers. However, as shown in Fig.~\ref{fig7} the electrical conductivity of T14\r{A} is slightly higher than that of T11\r{A} for the same temperature and carrier concentrations in both the $p$-type and $n$-type doping.
\subsubsection*{Electronic Thermal Conductivity}
\begin{figure}[!h]
    \centering
    (i)\includegraphics[width=0.655\textwidth]{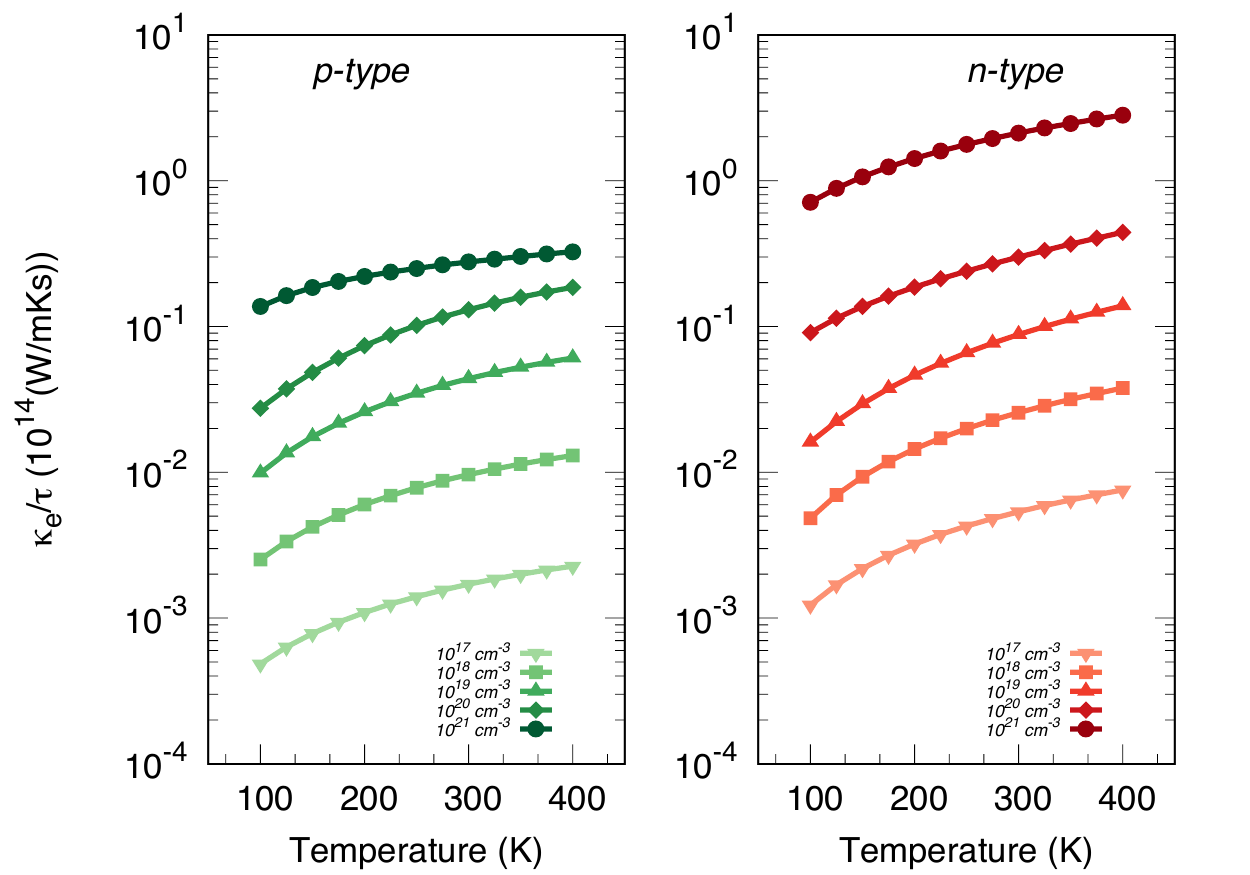}\\ 
    (ii)\includegraphics[width=0.655\textwidth]{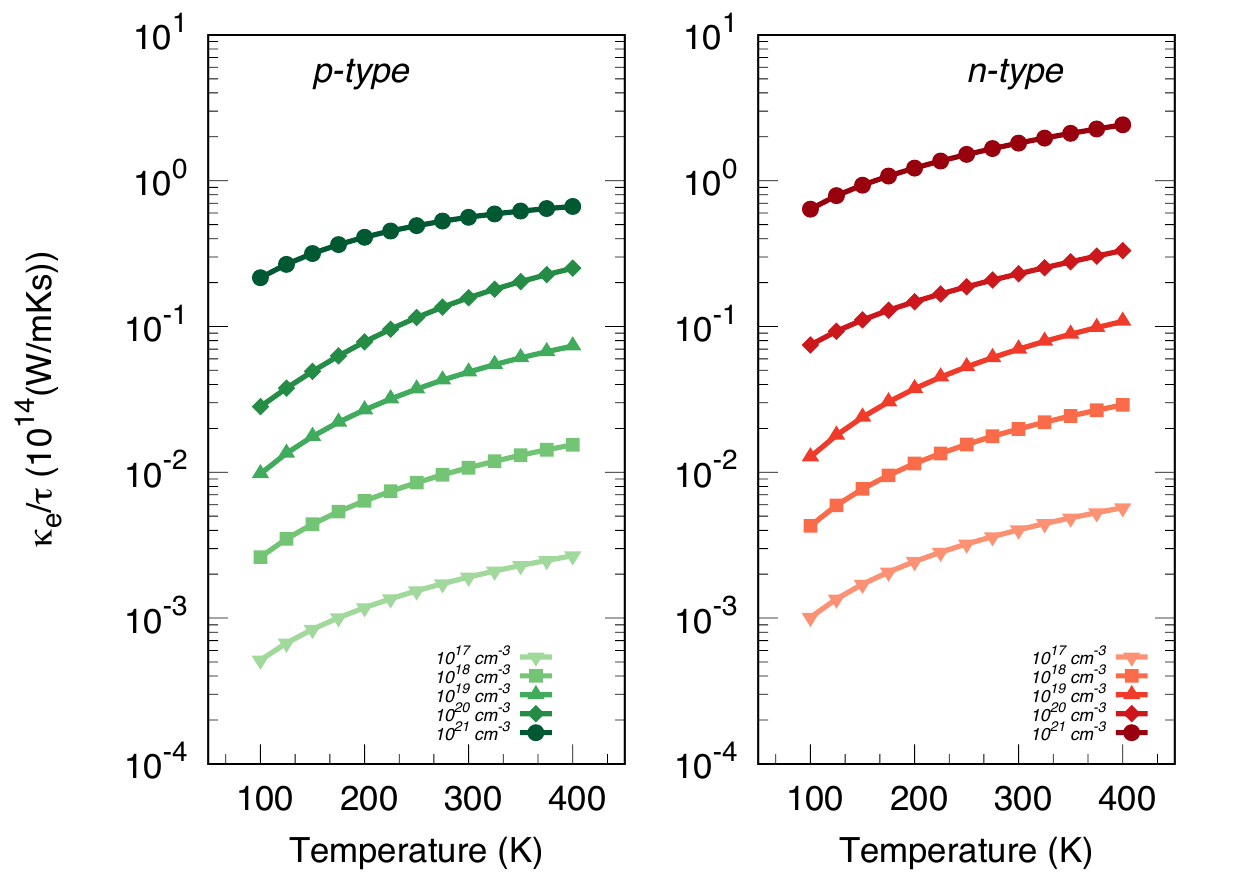}\\
    \caption{Averaged electronic thermal conductivity $\kappa_e/\tau$, as a function of temperature and carrier concentrations for T11\r{A}(i) and T14\r{A}(ii) models.}
    \label{fig8}
\end{figure}

The electronic component of the total thermal conductivity is given by the Wiedemann-Franz relation:
\begin{equation}
 \kappa_e=L\sigma T,\label{eqn7}
\end{equation}
where $L$ is the Lorenz number\cite{Snyder2008} and takes the value $L=2(k_B/q)^2$ for the non-degenerate case \cite{nolas2004}.
We evaluated the averaged electronic thermal conductivity $\kappa_{e}$ as a function of temperature and carrier concentration for T9\r{A} for both $p$-type and $n$-type doping using the values of $\tau$ computed from Sec.~\ref{Sec:RelaxTime}. The results are presented in Fig.~\ref{fig6}(ii). For $p$-type doping $\kappa_e$ decreases slowly with temperature as the carrier concentration increases, except at carrier concentration 10$^{20}$ cm$^{-3}$, where the magnitude of $\kappa_e$ initially increases up to 325 $K$ and then slightly decreases. In contrast, for the $n$-type $\kappa_e$ slightly increases (1\%) with temperature at carrier concentrations of $10^{18}$, $10^{19}$ and $10^{21}$ cm$^{-3}$, and decreases with temperature at carrier concentrations of $10^{17}$ and $10^{20}$ cm$^{-3}$. The average electronic thermal conductivity for the tobermorite 11\r{A} and 14\r{A} models was also evaluated and the plots are presented in Fig.~\ref{fig8}. 

It exhibits a monotonic increase with carrier concentration in the entire temperature range when the former increases from $10^{17}$ to $10^{21}$ cm$^{-3}$, for both $p$-type and $n$-type doping and for both T11\r{A} and T14\r{A}. This linear increase in $\kappa_e/\tau$ can easily be explained through the Wiedemann-Franz law, Eq.~(\ref{eqn7}), i.e. multiplying the decreasing values of $\sigma/\tau$ by the operating temperature, raises the values of $\kappa_e/\tau$. The electronic thermal conductivity exhibits a monotonic increase with temperature, the $n$-type values being an order of magnitude higher than those of the $p$-type doping over the entire carrier concentration range, for both T11\r{A} and T14\r{A}. 

\subsubsection*{Power Factor}
\begin{figure}[!h]
    \centering
    (i)\includegraphics[width=0.6\textwidth]{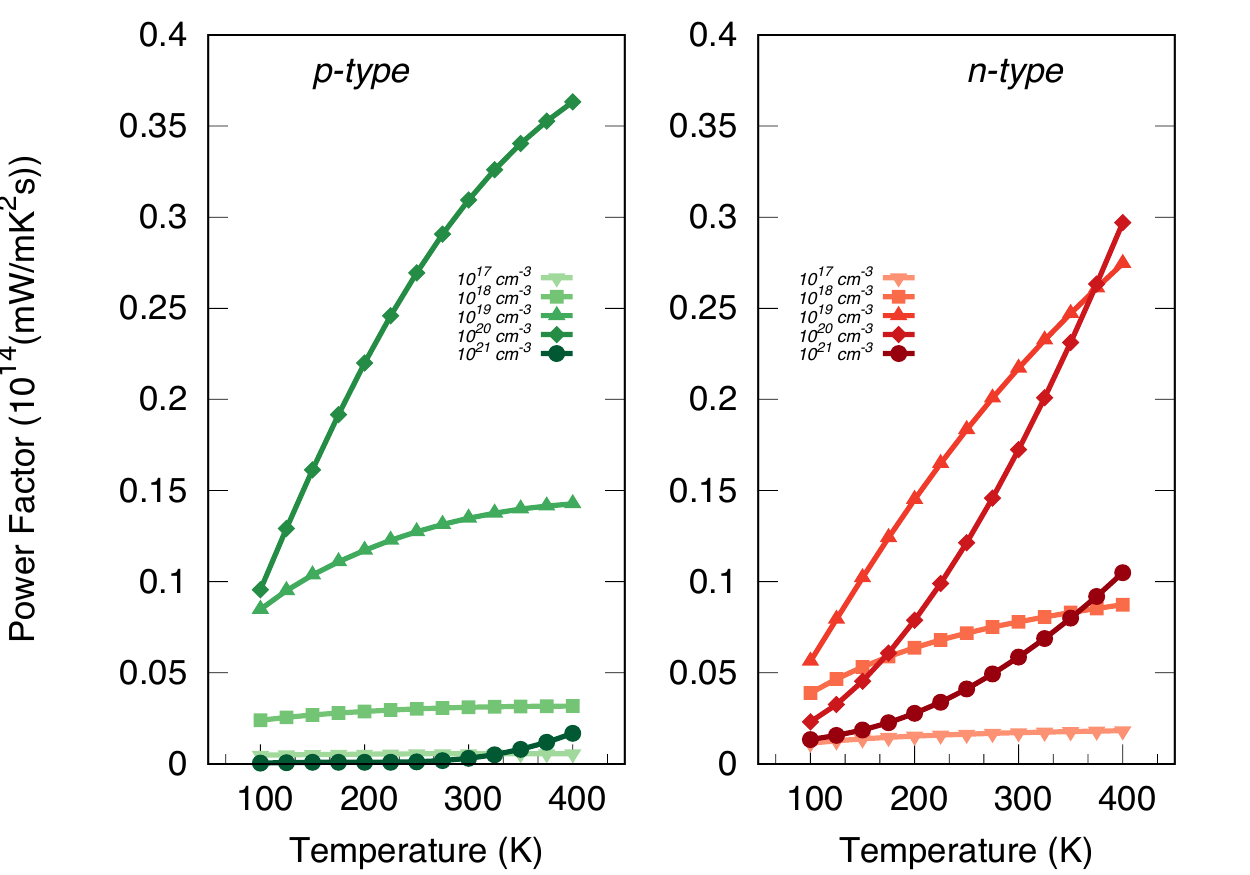}\\ 
    (ii)\includegraphics[width=0.6\textwidth]{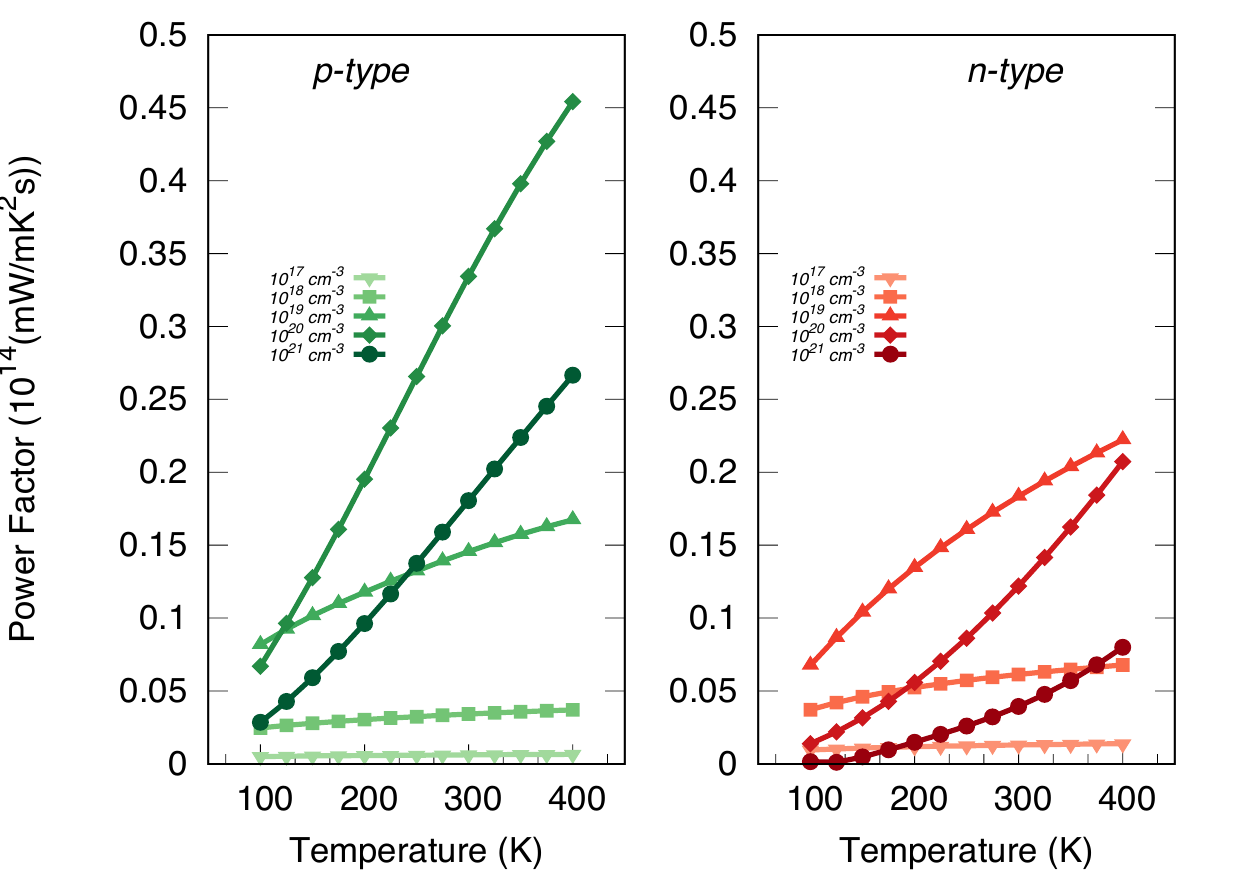}\\
    (iii)\includegraphics[width=0.6\textwidth]{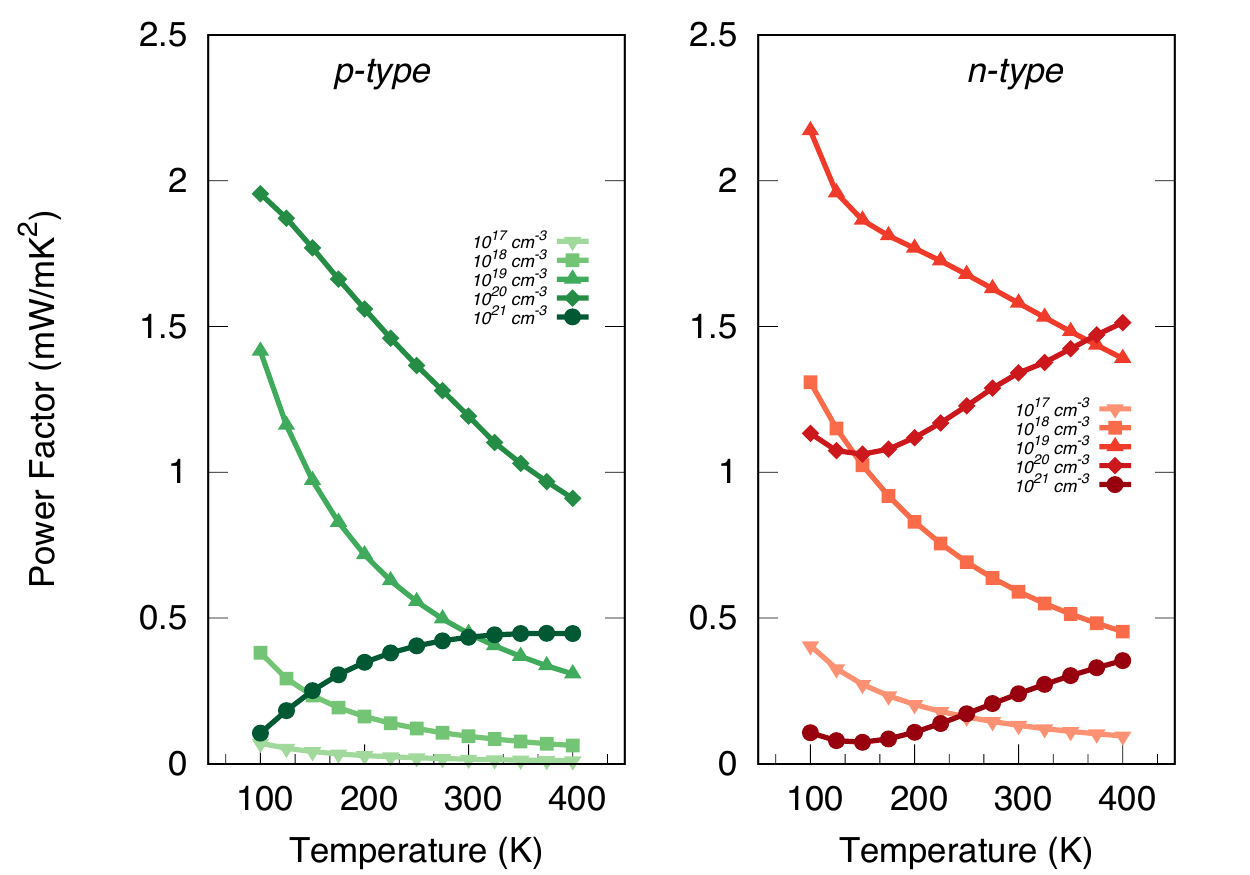}\\
    \caption{Calculated averaged Power factor $S^2\sigma/\tau$, as a function of temperature and carrier concentrations for T11\r{A}(i) and T14\r{A}(ii) models and power factor $S^2\sigma$ versus carrier concentration and temperature for T9\r{A}(iii) model.}
    \label{fig9}
\end{figure}

The power factor, $S^2\sigma/\tau$ was evaluated as a function of temperature and carrier concentration for the T11\r{A} and T14\r{A} models using the values of $S$ and $\sigma/\tau$ obtained from BoltzTraP. Results are presented in Fig.~\ref{fig9}. 
Both $p$-type and $n$-type power factors for T11\r{A} and T14\r{A} increase monotonically with temperature over the considered carrier concentration range. This is due to $S^2$ increasing faster than $\sigma/\tau$ decreases. However, at carrier concentrations of $10^{17}$ and $10^{21}$ cm$^{-3}$ in T11\r{A}, the power factor is increasing with temperature at a very negligible rate as compared to that of $n$-type doping, due to the almost zero $S$ and very low $\sigma/\tau$ values. A similar behaviour was observed in T14\r{A} at $10^{17}$ cm$^{-3}$. In $p$-type T11\r{A} (T14\r{A}) at $10^{20}$ cm$^{-3}$ concentration, the power factor rises to a maximum of $\approx 0.36\times 10^{14}$ mW/mK$^2$s ($\approx 0.45\times 10^{14}$ mW/mK$^2$s) at about 400 K. In $n$-type doping, the optimal power factors are slightly lower, hence suggesting that $p$-type doping holds the promise for a slightly better thermoelectric performance in tobermorite models. The temperature-dependent power factor $S^2\sigma$ versus carrier concentrations plot for T9\r{A} is also shown in Fig.~\ref{fig9}(iii). In both $n$-type and $p$-type power factor is seen to decrease as temperature increases except in higher carrier concentration values of $10^{20}$ and $10^{21}$ cm$^{-3}$. However, the power factor generally improved when the values of scattering time was added. 

\subsubsection*{Figure of merit ZT}
\begin{figure}[!h]
    \centering
    (i)\includegraphics[width=0.6\textwidth]{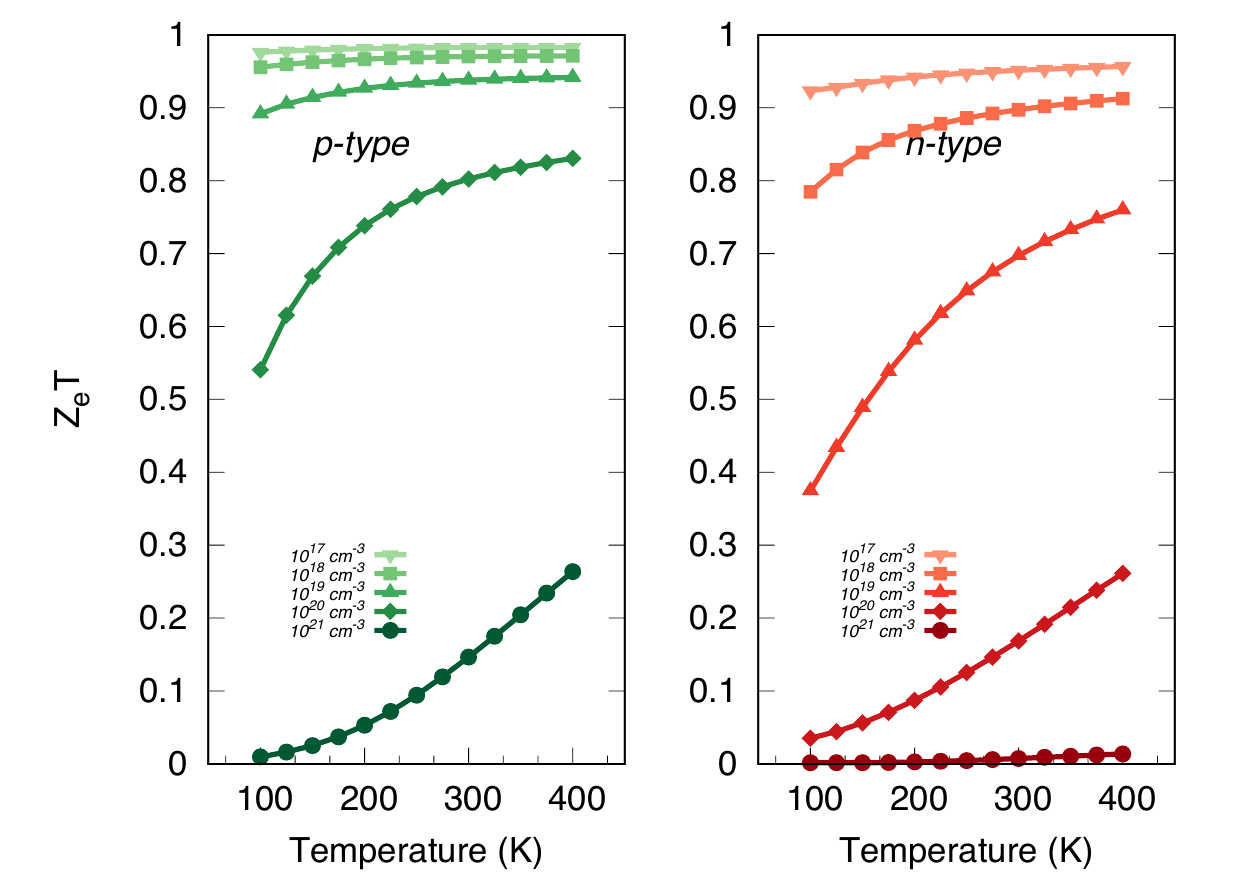}\\
    (ii)\includegraphics[width=0.6\textwidth]{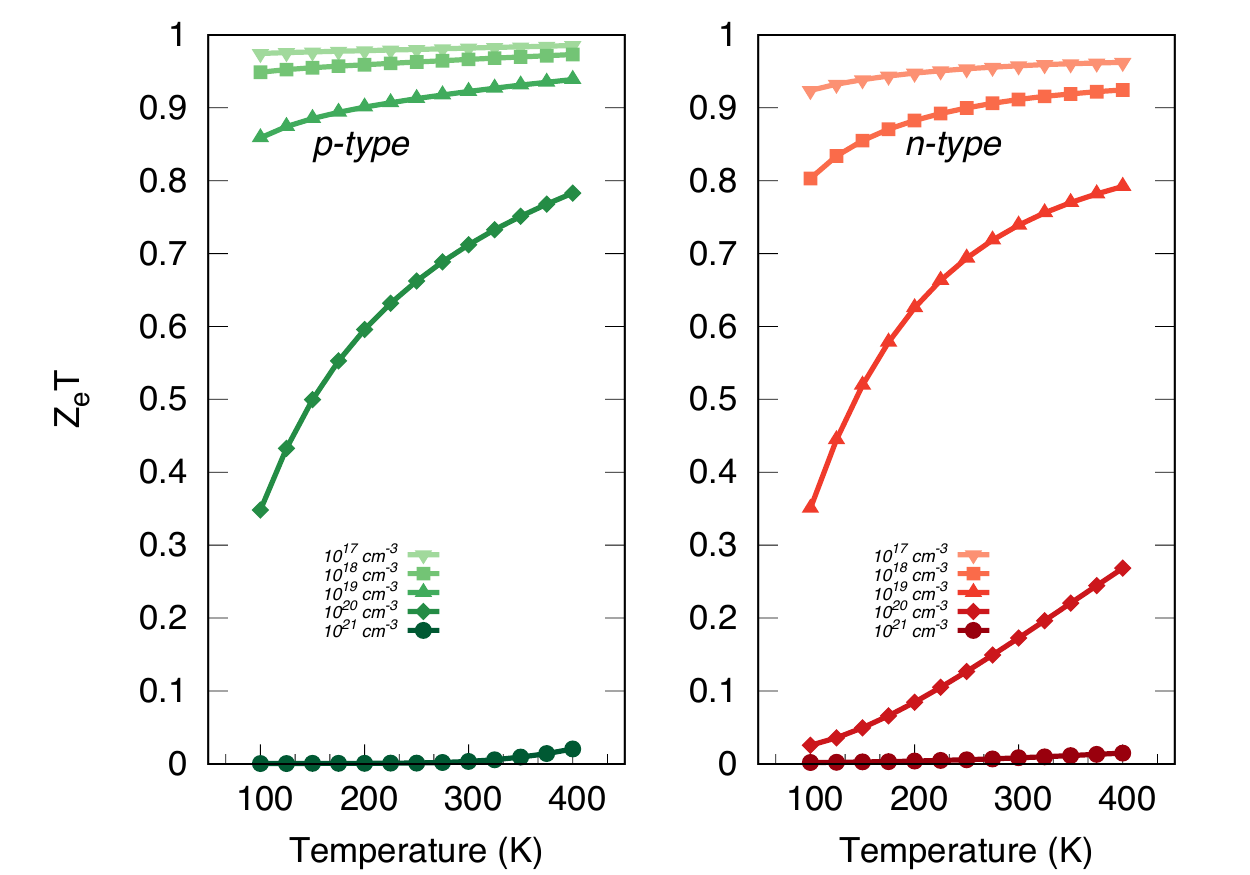}\\ 
    (iii)\includegraphics[width=0.6\textwidth]{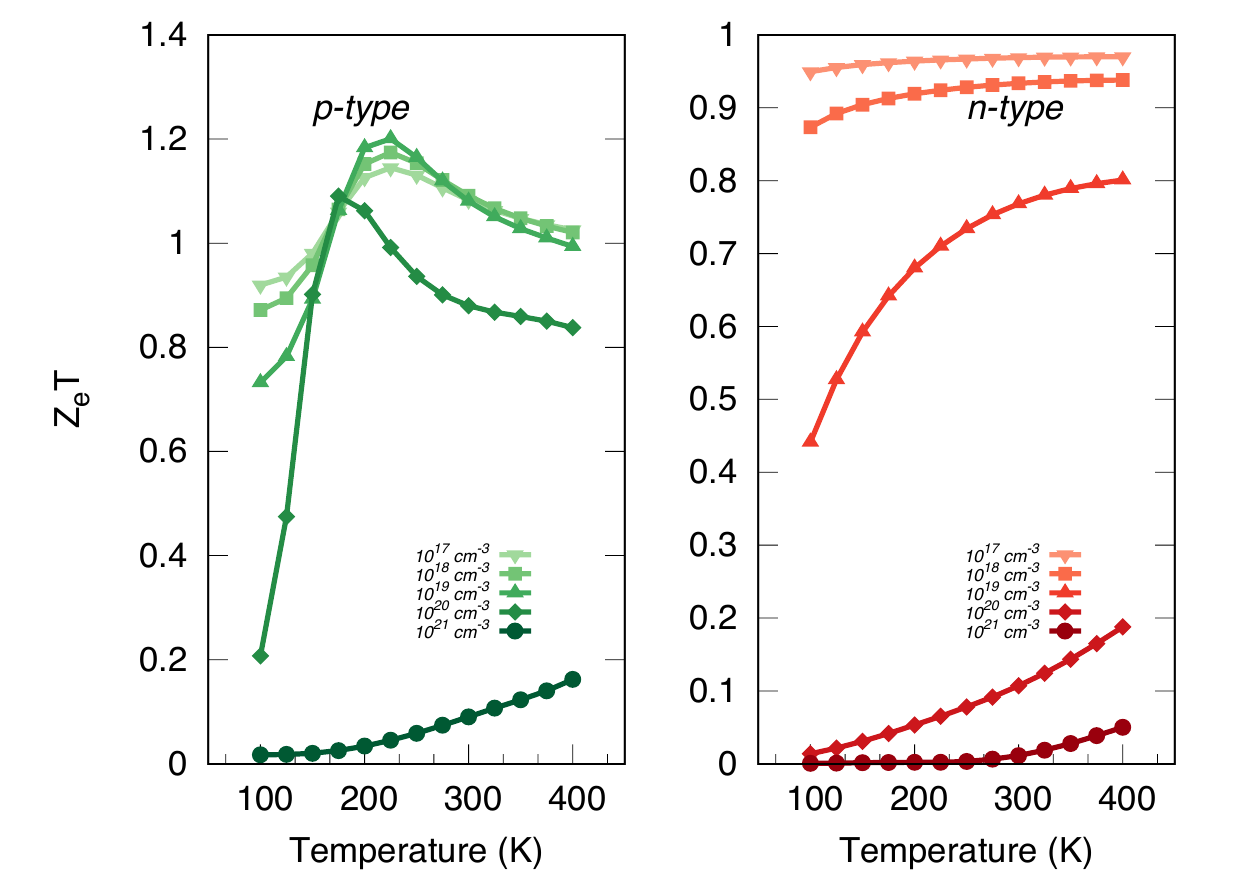}\\
    \caption{Calculated electronic $Z_eT$, as a function of temperature and carrier concentrations for T9\r{A}(i), T11\r{A}(ii), and T14\r{A}(iii) models. }
    \label{fig10}
\end{figure}

Using the calculated Seebeck coefficient and the values of $\sigma$ and $\kappa_e$, we calculated $Z_eT=S^2\sigma T/\kappa_e$, which is the figure of merit that ignores the lattice thermal conductivity $\kappa_l$, for $p$-type and $n$-type T9\r{A}. This is reported in Fig.~\ref{fig10}(i). The calculated $Z_eT$ values of $p$-type T9\r{A} are larger than those of $n$-type doping, because of the higher Seebeck coefficient (see Fig.~\ref{fig4}(a)). The calculated $Z_eT$ decreases for increasing carrier concentration, with the optimal values at $10^{17}$ cm$^{-3}$ in both $p$-type and $n$-type T9\r{A}. The maximum $Z_eT$ value at $10^{17}$ cm$^{-3}$ for $p$-type doping is $0.983$, which is constant from 275-400 K, while for $n$-type doping the value is 0.956 from 375-400 K. At 300 K, $Z_eT$ takes the value 0.951 for $n$-type doping at $10^{17}$ cm$^{-3}$.
In Fig.~\ref{fig10} we report the carrier concentration and temperature dependence of $Z_eT$ for $p$-type and $n$-type T11\r{A} and T14\r{A}. These were calculated using the Seebeck coefficient, electrical conductivity, $\sigma/\tau$ and electronic thermal conductivity, $\kappa_e/\tau$ obtained from BoltzTraP. Notice that the relaxation time $\tau$ is not needed to compute $Z_eT$, as it cancels out in the ratio $\sigma/\kappa_e=(\sigma/\tau)/(\kappa_e/\tau)$. From Fig.~\ref{fig10}(ii) (T11\r{A}) and \ref{fig10}(iii) (T14\r{A}), it is clear that $p$-type exhibits larger $Z_eT$ compared to $n$-type over the whole range of carrier concentrations. The optimal values of
$Z_eT$ slightly decrease with carrier concentration, but overall they remain close to 1, being optimal at $400$ K for $p$-type T11\r{A}. For $p$-type T14\r{A} the optimal $Z_eT$ values of 1.14, 1.17 and 1.20 correspond to a temperature of $225$ K and carrier concentrations from $10^{17}$ to $10^{19}$ cm$^{-3}$. For $n$-type doping, the highest values of $Z_eT$ are again slightly below 1 and decreasing mildly with concentration for both T11\r{A} and T14\r{A}. These results suggest that these tobermorite cement-based material would perform better as a $p$-type thermoelectric material in the low temperature regime. \\
To illustrate the potential thermoelectric performance of cement-based composites, their overall figure of merit, $ZT$, we use in Eq.~(\ref{eqn3}) the following representative values for the total thermal conductivity, $\kappa = \kappa_e + \kappa_l = 1.15$ W/mK \cite{Jani2022} and doping, $n = 10^{19}$ cm$^{-3}$.
Carrier densities as high as $1.626\cdot10^{18}$ cm$^{-3}$ have been reported in the literature.\cite{Wei2021}
Values of the total thermal conductivity in the interval $0.53$-$1.15$ W/mK have been reported for plain (\emph{i.e.}, with no additives) hydrated cement paste.\cite{Wei2018,Jani2022}.
We also fix the scattering rate, $\tau=10^{-14}$ s for tobermorite cement models T11\r{A} and T14\r{A}.

\begin{figure}[!h]
    \centering
    (i)\includegraphics[width=0.6\textwidth]{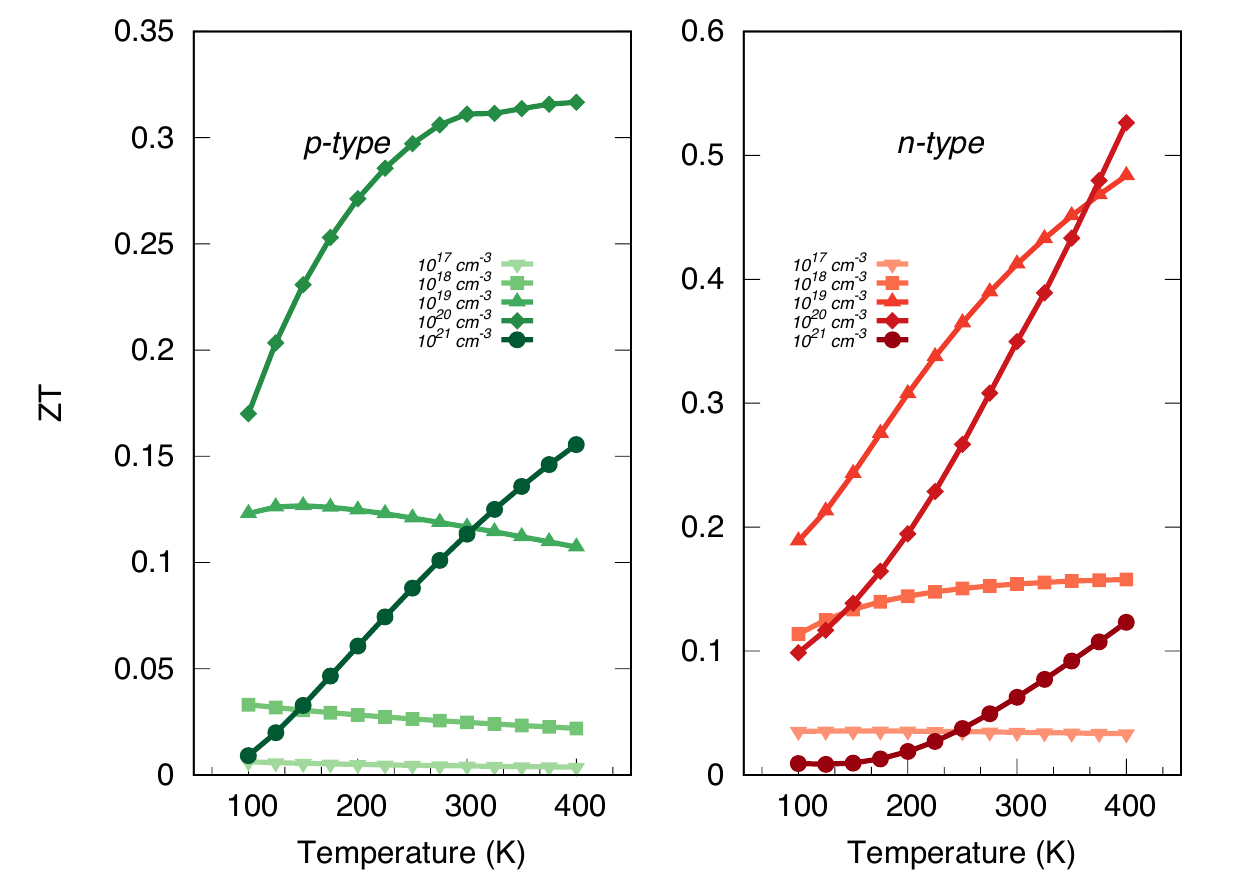}\\
    (ii)\includegraphics[width=0.6\textwidth]{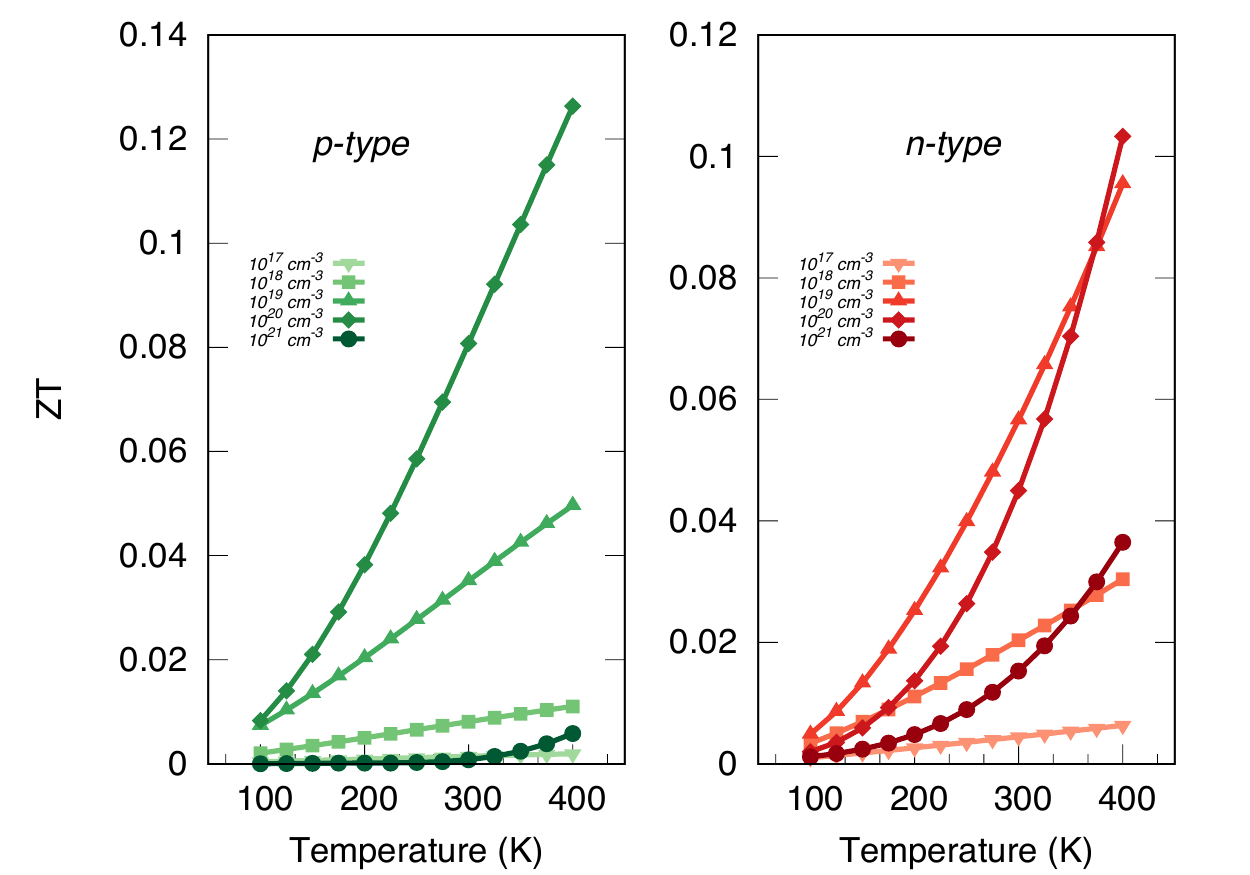}\\ 
    (iii)\includegraphics[width=0.6\textwidth]{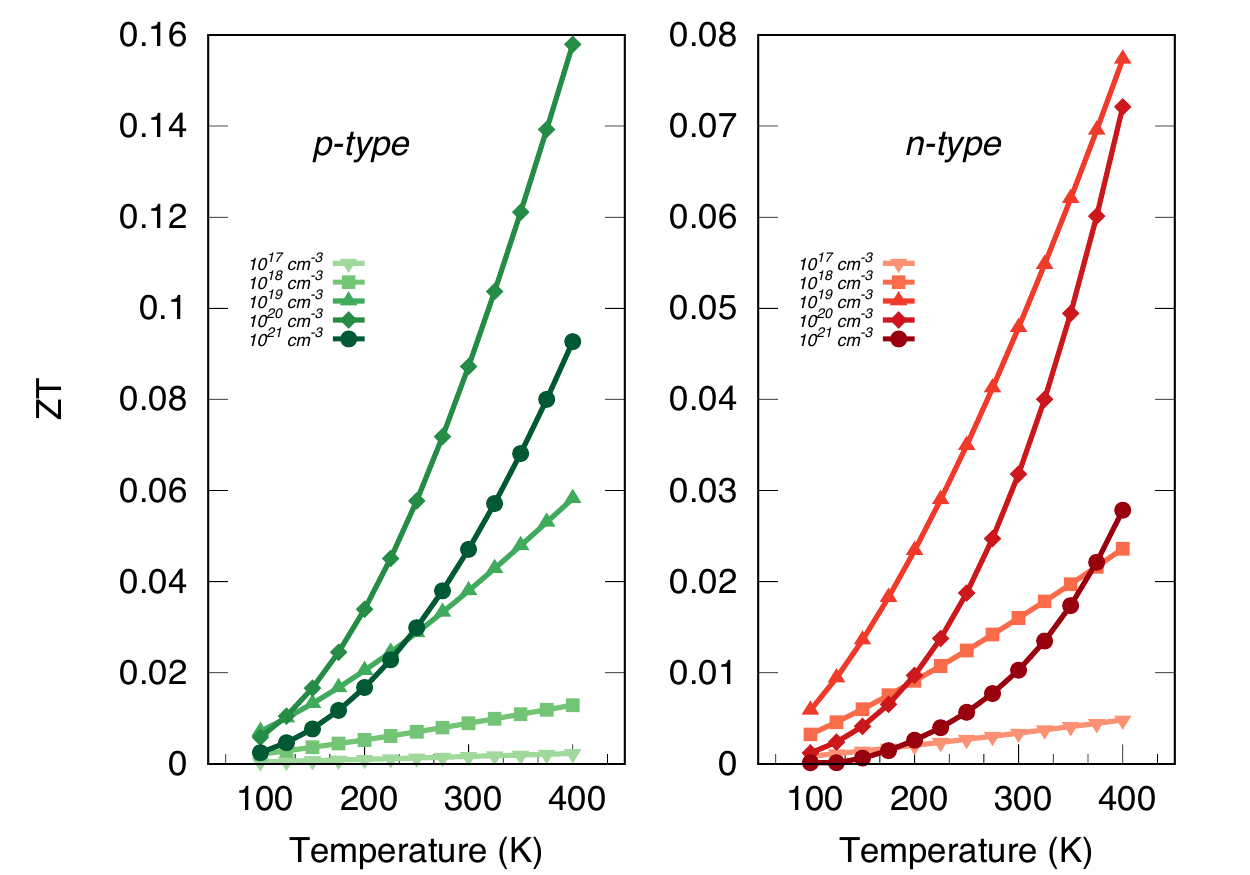}\\
    \caption{Calculated electronic $ZT$, as a function of temperature and carrier concentrations for T9\r{A}(i), T11\r{A}(ii), and T14\r{A}(iii) models.}
    \label{fig11}
\end{figure}

In Fig.~\ref{fig11}, based on our assumptions, our calculated results (ZT) for a typical plain hydrated cement paste show that the contribution of the lattice thermal conductivity, $\kappa_l$ is very significant and that the figure of merit is of the order $<0.1$ for lower carrier density $10^{17}$ to $10^{19}$ cm$^{-3}$  and a scaling factor $>0.1$ for higher carrier density. This means that the actual values of $ZT$ may be one tenth or slightly more of what we reported in the figures for the electronic figure of merit, $Z_eT$. From Fig. \ref{fig11} (i), it can also be seen that the inclusion of the scattering rate obtained via the simple parabolic band method, play a significant role as well in the determination of the figure of merit. Overall, our calculations indicate that $p$-type cement materials are the promising thermoelectric material.

\section{Conclusions}
We have, for the first time, employed first principles methods and the semi-classical Boltzmann theory to systematically investigate the electronic and transport properties of tobermorite minerals as an analog for cement for use in cement-based thermoelectrics. The calculated structural parameters obtained were seen to be comparable with the available experimental and DFT data. The electronic structure of these materials showed that they are direct band gap insulators, whose band gap (already mentioned in Table \ref{tab1}) decreases as the concentration of chemically bound water molecules increases. The calculated electronic relaxation time using a simple parabolic band approach showed a decreasing trend with increasing temperature, over the entire range of carrier concentrations. The electronic relaxation time  $\tau$ along the $a$ and $b$ crystallographic directions changes quite slowly as carrier concentration increases from $10^{17}$ to $10^{21}$ cm$^{-3}$,  because of the small difference in their longitudinal sound velocities. To the best of our knowledge there are no experimental data reported on transport properties of tobermorites, hence the results presented here are to be taken as predictions. However these results can also be affected by uncertainties especially the figure of merit $ZT$, which is based on the total thermal conductivity. As we tried to clarify these uncertainties based on the assumptions we made in Fig. \ref{fig11}, it can be seen that the contribution from lattice vibrations of these cement materials is essential and can not be neglected. We further saw from Fig. \ref{fig11} (i) that the contribution from the scattering time $\tau$ is appropriate from carrier concentration of $10^{19}$ to $10^{21}$ for both $p$-type and $n$-type materials but not much effective at carrier concentration of $10^{21}$ cm$^{-3}$. We found that the electronic transport properties of T9\r{A}, T11\r{A} and T14\r{A} favor the $p$-type thermoelectric behaviour. This is because of the presence of heavy, and thus flat, valence bands, which lead to a higher effective mass and hence a larger Seebeck coefficient. In contrast, the effective mass for electrons in the conduction band is an order of magnitude smaller, hence leading to smaller Seebeck coefficients in $n$-doped tobermorites. The small value of the electronic thermal conductivity suggests that the total thermal conductivity is mainly dominated by the intrinsic lattice thermal conductivity.
\\
Our study provides a blueprint for further computational studies of emerging cement composite materials of interest for thermoelectric applications, in particular those combined with carbon fibre . We have demonstrated from first principles how the Seebeck coefficient of a candidate material can be predicted \emph{in silico}, allowing for theoretical modifications to the material to be proposed and evaluated rapidly, in order to maximise the thermoelectric figure of merit.

\section*{Acknowledgements}
This research was supported through a US-Ireland grant funded by the Department for the Economy of Northern Ireland (DfE, USI 127). We are grateful for computational support from the UK Materials and Molecular Modelling Hub, which is partially funded by EPSRC (EP/P020194 and EP/T022213), for which access was obtained via the UKCP consortium and funded by EPSRC grant ref EP/P022561/1. JK was also supported by the Beatriz Galindo Program (BEAGAL18/00130) from the Ministerio de Educación y Formación Profesional of Spain, and by the Comunidad de Madrid through the Convenio Plurianual with Universidad Politécnica de Madrid in its line of action Apoyo a la realización de proyectos de I+D para investigadores Beatriz Galindo, within the framework of V PRICIT (V Plan Regional de Investigación Científica e Innovación Tecnológica)
\bibliographystyle{apsrev4-2}
\bibliography{ThermoConc-paper1}
\end{document}